\documentclass[12pt]{spieman}  
\usepackage{amsmath,amsfonts,amssymb}
\usepackage{graphicx}
\usepackage{setspace}
\usepackage{tocloft}

\usepackage{longtable}
\usepackage{gensymb}
\usepackage{color,soul}
\usepackage{lineno}

\title{Developing an error budget for the nonlinear curvature wavefront sensor}

\author[a,*]{Sam Potier}
\author[a]{Justin Crepp}
\author[a]{Stanimir Letchev}
\affil[a]{University of Notre Dame, Department of Physics and Astronomy, 225 Nieuwland Science Hall, Notre Dame, IN 46556, USA}

\cftpagenumbersoff{figure}
\cftpagenumbersoff{table} 
\begin{document} 
\maketitle

\begin{abstract}
Consistent operation of adaptive optics (AO) systems requires the use of a wavefront sensor (WFS) with high sensitivity and low noise. The nonlinear curvature WFS (nlCWFS) has been shown both in simulations and lab experiments to be more sensitive than the industry-standard Shack-Hartmann WFS (SHWFS), but its noise characteristics have yet to be thoroughly explored. In this paper, we develop a spatial domain wavefront error budget for the nlCWFS that includes common sources of noise that introduce uncertainty into the reconstruction process (photon noise, finite bit depth, read noise, vibrations, non-common-path errors, servo lag, etc.). We find that the nlCWFS can out-perform the SHWFS in a variety of environmental conditions, and that the primary challenge involves overcoming speed limitations related to the wavefront reconstructor. The results of this work may be used to inform the design of nlCWFS systems for a broad range of AO applications.
\end{abstract}


{\noindent \footnotesize\textbf{*}Sam Potier,  \linkable{spotier@nd.edu} }

\begin{spacing}{2}   

\section{Introduction}\label{sec:intro}
A wavefront sensor's (WFS) ability to measure phase and amplitude distortions governs the performance of active and adaptive optics (AO) systems. An ideal WFS is inherently sensitive, has low systematic errors and a large dynamic range, and can realize fast reconstruction speeds. For example, when driven by the need to correct for image blurring caused by atmospheric turbulence, these requirements demand an operating frequency of many kilo-Hertz using low illumination levels \cite{Davies:12}.

Many WFS designs have been developed to compensate for Earth's turbulent atmosphere, including the Shack-Hartmann \cite{Platt:01}, pyramid \cite{Ragazzoni:96}, curvature \cite{Roddier:88}, interferometric \cite{Angel:94}, and others. The industry-standard Shack-Hartmann WFS (SHWFS) uses a lenslet array to produce a field of image plane spots whose positions correspond to local tip/tilt phase estimates. Although an optically robust design, this approach lacks sensitivity at low spatial frequencies compared to a theoretically ideal sensor \cite{Guyon:10}, and increasing spatial sampling for a flux limited application only exacerbates the problem \cite{Mateen:15}. Further, the SHWFS is also susceptible to irradiance fade in strong scintillation environments \cite{Watnik:18,Crepp:20}.

\begin{figure}
    \centering
    \includegraphics[width=0.6\textwidth]{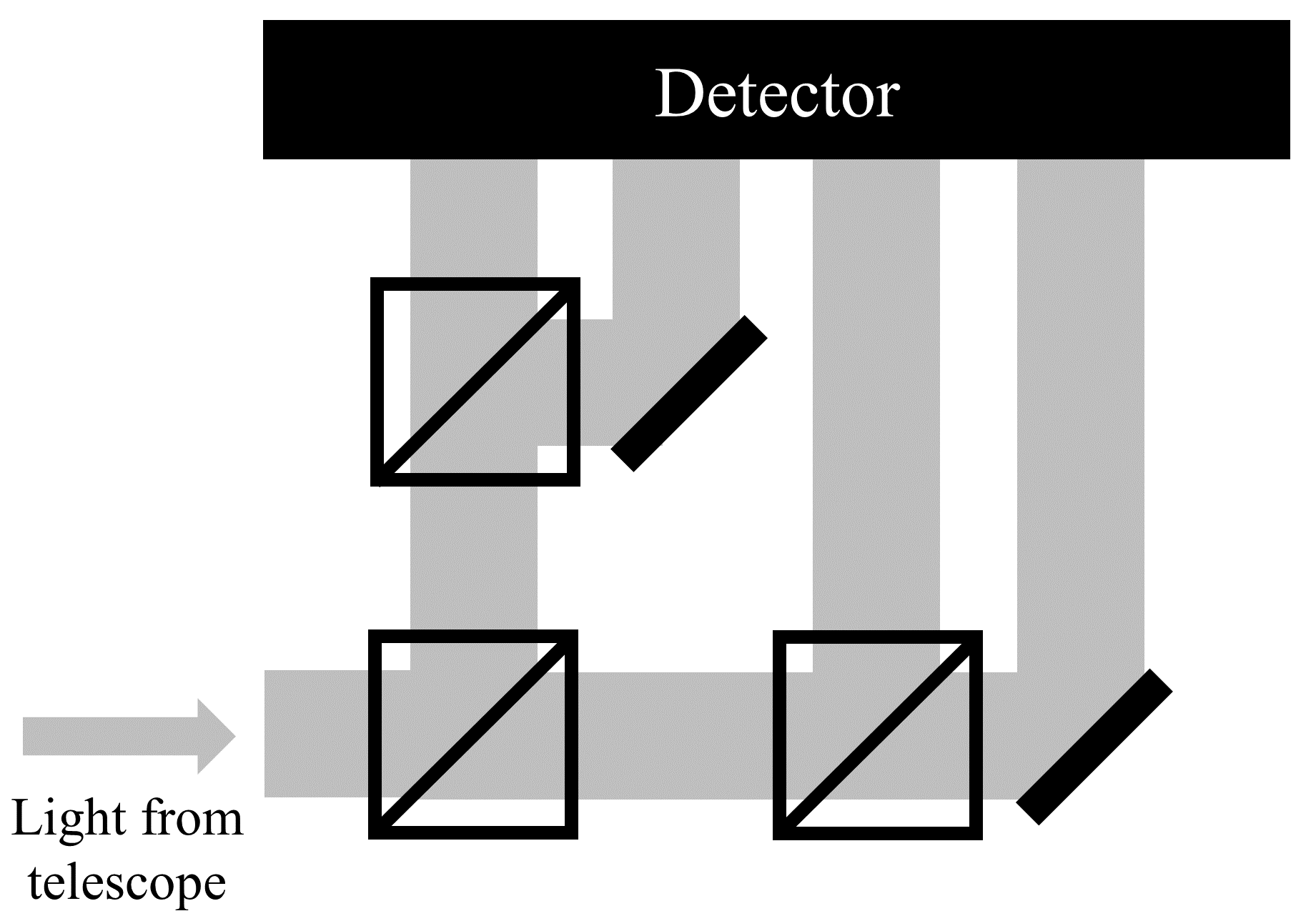}
    \caption{Schematic of the nlCWFS. Beam-splitting optics divide the light received from the telescope into four channels, enabling the capture of phase diversity in the form of path length differences.}
    \label{fig:nlCWFS_diagram}
\end{figure}

In this study, we investigate replacing the SHWFS with a nonlinear curvature WFS (nlCWFS) to improve the reconstruction accuracy of remote sensing observations. Originally developed for astronomy, the nlCWFS samples a propagating beam at (typically) four locations along the optical axis that are displaced from the pupil. Intensity measurements from these spatial samples (hereafter `measurement planes'), which provide phase diversity in the form of path length differences, are sent to a modified version of the Gerchberg-Saxton algorithm to reconstruct wavefront phase and amplitude. Figure \ref{fig:nlCWFS_diagram} shows a schematic of the nlCWFS. One of the most photon-efficient WFS designs, reliance on internal beam interference allows the nlCWFS to be more sensitive than the SHWFS over a broad range of spatial frequencies \cite{Guyon:10}, a result that has been corroborated by both numerical studies \cite{Letchev:22} and lab experiments \cite{Crass:14,Mateen:15,Crepp:20}. 




Most investigations of the nlCWFS to date have assumed operation of an ideal, noiseless sensor. We aim to compare the wavefront reconstruction accuracy of the nlCWFS to comparable SHWFS designs in the presence of common sources of noise and error encountered in practice. If the nlCWFS is to serve as an effective alternative to the SHWFS for real-world applications, then its performance must be robust to environmental conditions as validated through a (near-) comprehensive error budget in both simulations and lab experiments \cite{Ahn:23}. This paper describes the first step in the process by testing the sensitivity of a non-ideal nlCWFS using numerical modeling. 


In Sec. \ref{sec:methods}, we describe the numerical methods used to simulate nlCWFS and SHWFS device modules and common environmental and systematic error sources. In Sec. \ref{sec:results}, we present results for individual physical effects that can contribute to residual wavefront error (WFE) using uncorrected, spatial domain measurements. In Sec. \ref{sec:budget}, we develop a representative WFE budget that compares the nlCWFS to an equivalent SHWFS. In Sec. \ref{sec:conclusions}, we present the conclusions of the study.

\section{Methods}\label{sec:methods}
We quantify wavefront reconstruction errors relative to an ideal measurement device when perturbations to the optical system, detector, or reconstruction process are incurred. Results are cast in terms of WFE to create a wavefront error budget (phase only). Individual noise and error sources that are simulated include: photon noise, finite bit depth, detector read noise, detector vibrations, non-common-path aberrations, and servo lag. We assume all error sources to be uncorrelated and thus simulate them independently. We have also developed simulations involving scintillation in low flux regimes but this will be the subject of a forthcoming article \cite{Potier:23}.

\subsection{Model Description}\label{sec:model}
\begin{figure}
    \centering
    \includegraphics[width=\textwidth]{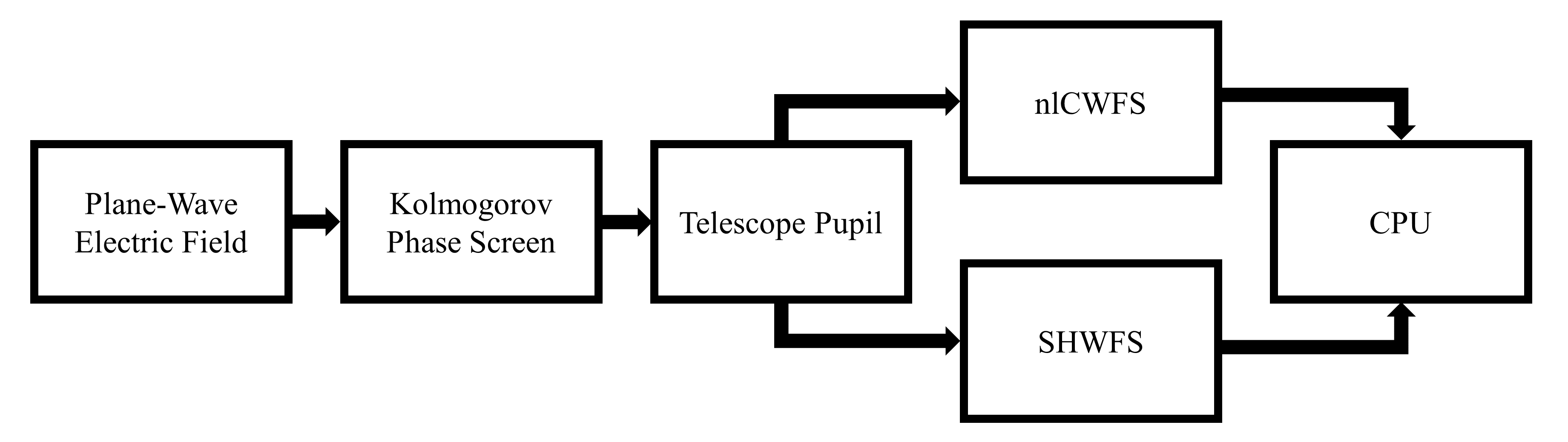}
    \caption{Block diagram of the simulations used to compare the SHWFS and nlCWFS.}
    \label{fig:Block_diagram}
\end{figure}

Numerical simulations in Matlab were employed using the \emph{WaveProp} and \emph{AOTools} toolboxes to model atmospheric turbulence, optical propagation, and wavefront sensing \cite{Brennan:16,Brennan:17}. Figure \ref{fig:Block_diagram} shows a block diagram of the simulations used in this study. We create an input electric field as a coherent plane-wave with wavelength $\lambda$ = 532nm. This wavelength was selected to help inform on-going lab experiments of the nlCWFS. A previous study at Notre Dame has studied the impact of broadband illumination on nlCWFS performance \cite{Letchev:22}. The complex field is distorted by including phase aberrations that follow a Kolmogorov power spectrum. Different phase screens are made using WaveProp's \emph{KolmogPhzScreen} class, which uses a spatial spectral index of $\phi$ = -11/3 \cite{Tatarski:61}. 

Thirty distinct tip-and-tilt-removed instances of atmospheric turbulence are simulated to statistically quantify WFS performance. We assume a $D_{\textrm{tel}}=1$ m telescope with a primary mirror that has a perfect surface. We do not include a secondary mirror nor central obstruction. The input electric field, telescope pupil, and subsequent optical propagations are modeled using a 896$\times$896 array to maintain a turbulence sampling greater than 9 grid points per $r_0$, the Fried parameter \cite{Fried:66}. For wavefront algorithmic errors, $r_0$ is varied between 2 - 20 cm (Sec. \ref{sec:AE_methods}); $r_0$ is set to 10 cm for all other simulations (Secs. \ref{sec:PN_methods} - \ref{sec:HFNCP_methods}). 

\subsection{Wavefront Sensing}\label{sec:sensing}
Following application of the telescope to create a pupil, the beam is then passed to the WFS (either the SHWFS or nlCWFS). To isolate physical effects in the spatial domain of the WFS, we consider operation using an uncorrected beam for sensing only, i.e. no deformable mirror in the system. A future article will explore closed-loop control in detail and focus on time domain effects.

We simulate multiple configurations of the SHWFS and nlCWFS to span a representative tradespace between WFS reconstruction accuracy and robustness to various sources of noise and error. WFS spatial sampling parameters are chosen to match previous \emph{WaveProp} and \emph{AOTools} WFS studies \cite{Mateen:15,Spencer:20}, the lab setup at the University of Notre Dame \cite{Crepp:20}, and a concurrent study into SHWFS and nlCWFS performance in the presence of scintillation \cite{Potier:23}. 

Using \emph{WaveProp}'s \emph{HartmannSensor} class, the SHWFS is modeled with lenslet arrays of 32$\times$32 and 64$\times$64 square lenslets across the pupil (referred to as SHWFS-32 and SHWFS-64). An 8$\times$8 array of detector pixels is used to sample the light behind each lenslet. We assume that the SHWFS and nlCWFS detectors have large regions with well-behaved pixels; we do not model hot pixels or dead pixels. The SHWFS lenslet pitch ($d$) and focal length ($f$) are set to provide a lenslet field-of-view of $5.65~\lambda/d$. Light passing through the lenslet array creates a spot-field that we use to calculate local phase gradients weighted by intensity. We then perform a wavefront reconstruction by sending the gradients to \emph{AOTools}' least-squares reconstructor \cite{Brennan:16}. We assume a throughput of 100\% for both SHWFS configurations, ignoring the marginal throughput loss that results from propagation through the SHWFS's lenslets.


We likewise model two nlCWFS configurations with spatial samplings of 32$\times$32 and 64$\times$64 across the beam as reconstructed at the pupil plane (referred to as nlCWFS-32 and nlCWFS-64). Assuming a constant beam size, varying the number of spatial samples per $r_0$ governs the accuracy of the nlCWFS reconstruction process. We find that the nlCWFS requires more than 2.5 samples per $r_0$ to reliably reconstruct wavefront phase information, maintaining close agreement with concurrent nlCWFS studies \cite{Letchev:22}. The default $r_0$ of 10 cm and nlCWFS beam size of 0.5 mm correspond to 3.2 and 6.4 samples per $r_0$ for the nlCWFS-32 and nlCWFS-64, respectively.


Sensitivity of the nlCWFS to certain spatial frequencies is further influenced by the path length differences between measurement planes. We simulate measurement plane positions of $\pm$1 cm and $\pm$4 cm as measured from the pupil. These distances have been found empirically to be large enough to provide the nlCWFS with sensitivity to low spatial frequencies, which dominate the aberration power spectrum, yet small enough to minimize losses in signal-to-noise ratio (SNR) from diffraction \cite{Letchev:22}. We reconstruct the wavefront phase (and amplitude) using a modified version of the Gerchberg-Saxton algorithm to accommodate four planes. The resulting $2 \pi$ ambiguous phase is then unwrapped using 
\emph{WaveProp}'s \emph{Sphase} phase unwrapping function \cite{Brennan:17}. We simulate a throughput of 65.9\% for both nlCWFS configurations, to account for signal losses that have been measured in lab experiments.



\begin{figure}
    \centering
    \includegraphics[width=\textwidth]{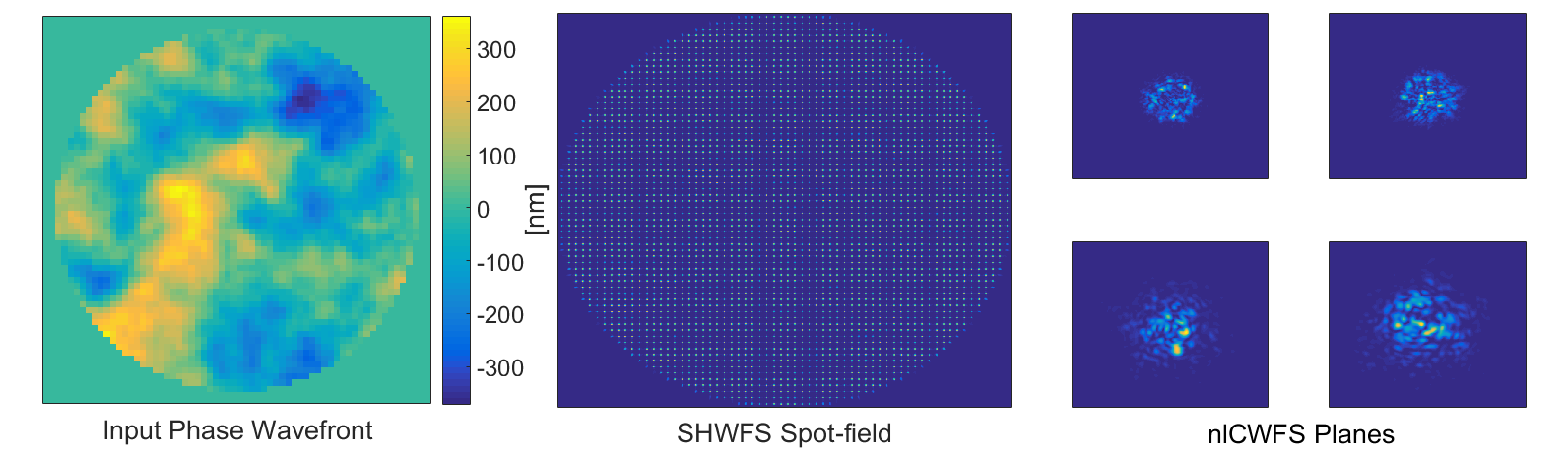}
    \caption{Example Kolmogorov wavefront phase aberration (left), the corresponding SHWFS spot-field (middle), and nlCWFS measurement planes (right).}
    \label{fig:wfs_samples}
\end{figure}

\begin{figure}
    \centering
    \includegraphics[width=0.98\textwidth]{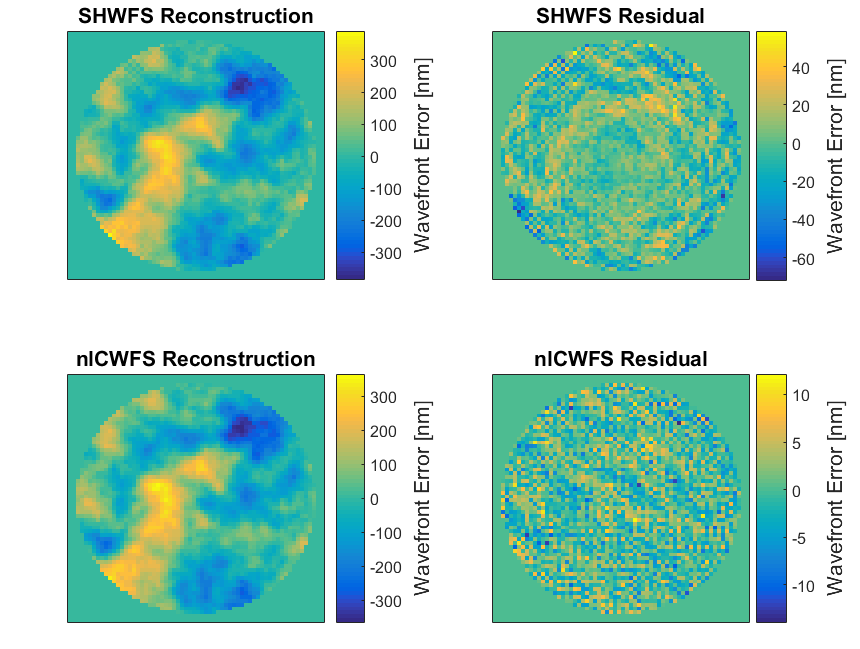}
    \caption{Using the same Kolmogorov phase aberration as in Fig. \ref{fig:wfs_samples}, the SHWFS (top-left) and nlCWFS (bottom-left) both produce reliable wavefront reconstructions. For a given flux level, the nlCWFS generally produces smaller residuals than the SHWFS (right panels). See text for discussion.}
    \label{fig:recon_example}
\end{figure}

Figure \ref{fig:wfs_samples} displays a representative Kolmogorov wavefront realization and the corresponding SHWFS spot-field and nlCWFS measurement planes. Figure \ref{fig:recon_example} presents SHWFS and nlCWFS reconstructions and their difference from the input wavefront. Qualitatively, both the SHWFS and nlCWFS reconstructions match well with the input wavefront (Fig. \ref{fig:wfs_samples}). However, the nlCWFS produces a more accurate wavefront estimate and, thus, a smaller residual (Fig. \ref{fig:recon_example}).

\subsection{Photon Noise}
\label{sec:PN_methods}
The arrival-time statistics of light quanta from lasers, stars, and other illumination sources place a fundamental limitation on WFS performance \cite{Guyon:05}. For short integration times, comparable to changes in atmospheric conditions, repeated flux measurements will yield variable results. We simulate the random nature of photon noise using methods provided by the \emph{WaveProp} toolbox, which uses Poisson statistics \cite{Mandel:59}. The number of photo-electrons in each pixel for a given exposure is quantified based on the radiation intensity. A perfect detector with quantum efficiency of 100\% and gain of unity is assumed. 

\subsection{Systematic Errors}\label{sec:systematics}

\subsubsection{Algorithmic Errors}\label{sec:AE_methods}
The nlCWFS and SHWFS employ different methods of wavefront reconstruction. Accuracy is ultimately limited by the algorithms themselves. We quantify reconstruction errors caused by algorithmic inaccuracies by comparing input wavefront phase aberrations to reconstructions in the absence of all other error sources.

\subsubsection{Finite Bit Depth}\label{sec:FBDD_methods}
The digitization process used by detectors introduces discrete numerical errors into measured intensity values that can impact the precision of wavefront reconstruction. While many modern cameras can transmit a reasonably large bit depth ($b$), there exists a trade-off between information accuracy and the transmission speeds needed by AO systems. We study how $b=8$ (256 counts) through $b=16$ (65,536 counts) digitization errors can influence the precision of the SHWFS and nlCWFS in 1-bit increments.

\subsubsection{Detector Read Noise}
\label{sec:RN_methods}

\begin{figure}
    \centering
    \includegraphics[width=\textwidth]{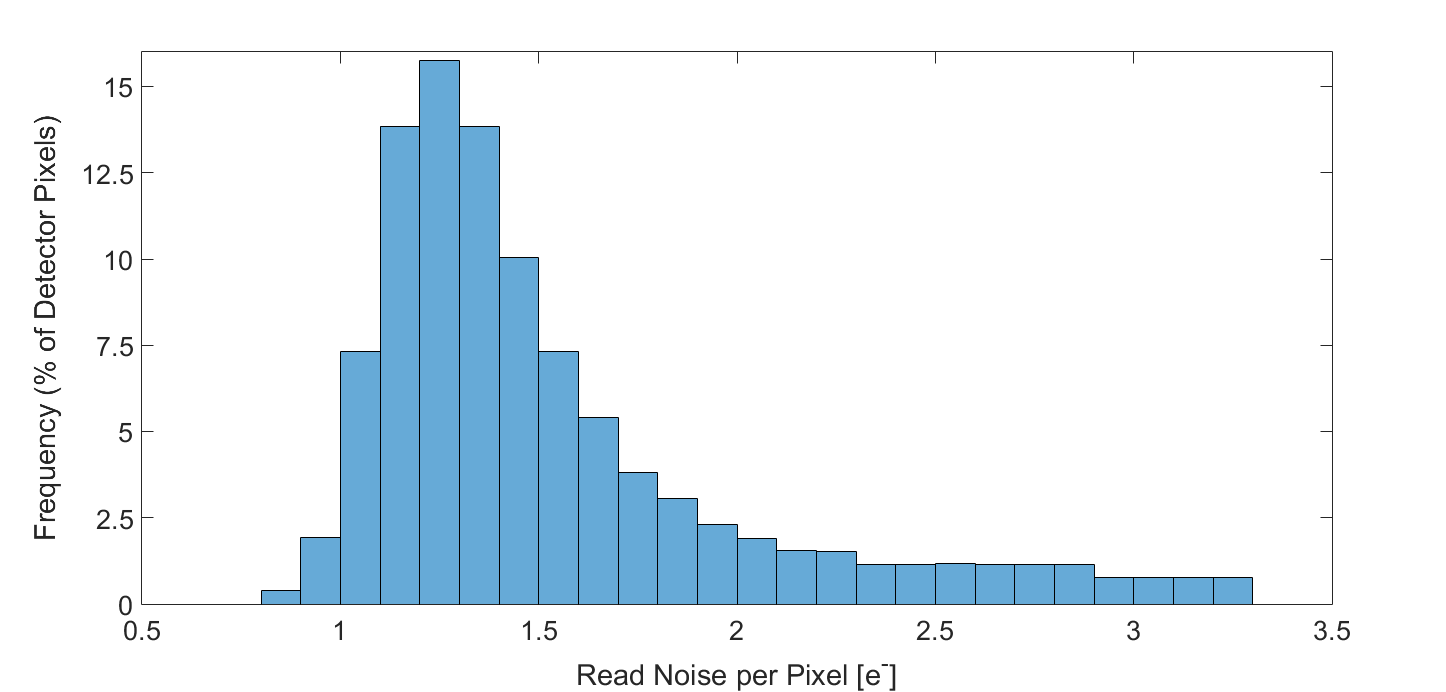}
    \caption{Representative sCMOS read noise distribution with median of 1.4 photo-electrons.}
    \label{fig:histo}
\end{figure}

We simulate read-out noise at each detector pixel using a random draw from the representative scientific Complementary Metal-Oxide Semiconductor (sCMOS) detector read noise histogram shown in Figure \ref{fig:histo} \cite{Andor:NA}. Unlike CCD detectors, the read noise distribution of a sCMOS detector is defined by its median, not its RMS. The sCMOS read noise distribution shown in Fig. \ref{fig:histo} has a median of 1.4 photo-electrons, which is similar to the device used for on-going experiments of the nlCWFS at Notre Dame \cite{Crepp:20}. We simulate different levels of read noise by varying the median read noise value. 

We study the impact of read noise with optimal pedestal subtraction on WFS reconstruction accuracy. Pedestal subtraction, defined as a uniform deduction across a WFS intensity sample following readout from the detector, helps to minimize noise that the SHWFS would otherwise include in its centroid calculations or the nlCWFS would interpret as diffracted light in its measurement planes. Optimal pedestal subtraction reduces pixel count variance while minimizing signal loss. We have found optimal pedestal subtraction to improve reconstruction results for both sensors and, thus, include it in all simulations of read noise.

To simulate pedestal subtraction, we apply a threshold to the images returned by each WFS by subtracting a pedestal value, $T_{e^-}$, from the entire image and setting any resulting negative values to zero. We define $T_{e^-}$ as a coefficient, $C_{e^-}$, multiplied by the read noise median $M_{e^-}$:

\begin{equation}
    T_{e^-}=C_{e^-} \; M_{e^-}.
\end{equation}


\noindent This pedestal subtraction technique is similar to that of Schlawin et al. 2020, who developed a noise reduction method for James Webb Space Telescope NIRCam measurements \cite{Schlawin:20}. For each row on the detector, Schlawin et al. 2020 remove the median pixel count of a non-illuminated portion of the detector from all pixels in the row. This matches the pedestal subtraction used in this study when $C_{e^-}$ is set to 1. We do not simulate row-by-row read noise correlation and, thus, apply pedestal subtraction evenly across the full detector array.

To determine the optimal $C_{e^-}$ values for this study, we simulated read noise with pedestal subtraction using a range of $C_{e^-}$ values and calculated the resulting SHWFS and nlCWFS RMS WFE. For each $M_{e^-}$, we selected the $C_{e^-}$ value that generated the smallest RMS WFE as optimal. For the nlCWFS, we find the optimal $C_{e^-}$ to be between 1.0 and 1.1 across all simulated $M_{e^-}$. Thus, the optimal pedestal for subtraction from each nlCWFS measurement plane sample has magnitude approximately equal to the detector read noise median. For the SHWFS, we find the optimal $C_{e^-}$ ranges from 0.8 to 6.3 depending on the value of $M_{e^-}$.



\subsubsection{Detector Vibrations}\label{sec:VIB_methods}
To prevent overheating from electronics, high-speed detectors often require active or passive cooling that can cause vibrations. For example, cryogenic cooling was found to cause vibrations at 60 Hz in the Gemini Planet Imager \cite{palacios:21}. We assume the detector vibrations occur on timescales longer than the WFS integration time and, thus, translate the detector and its connected optics off of the optical axis. For the nlCWFS, we model vibrations by translating the beam across the detector surface. For the SHWFS, we model vibrations by translating the beam across the lenslet array, which we assume to be secured to the detector. We do not simulate beam tracking, and detector regions-of-interest are assumed to be static. We measure vibration amplitude in terms of the WFS beam diameter ($D$) to normalize across the different beam diameters used for SHWFS (4 mm) and nlCWFS (0.5 mm).


\subsubsection{Non-Common-Path (NCP) Aberrations in the nlCWFS} \label{sec:NCP_methods}

The nlCWFS separates the incoming beam into multiple channels with distinct optical paths to create phase diversity that informs the reconstruction process. Wavefront distortions unique to each channel (non-common-path aberrations) reduce the correlation between the propagating electric fields. When supplied to the nlCWFS's reconstruction algorithm, sufficiently uncorrelated measurement plane samples introduce additional computation time and can prohibit an accurate wavefront estimate.


\paragraph{Differential Tip/Tilt}
\label{sec:DT_methods}

Equipment misalignments unique to each nlCWFS channel can introduce differential tip/tilt by changing the Poynting vector of each channel's electric field. The nlCWFS's reconstruction algorithm does not differentiate between channel-specific and global tilt, meaning that differential tip/tilt can limit the accuracy of the sensor's phase estimates. We simulate differential tip/tilt by applying a plane of tip and tilt with random direction but uniform angular magnitude, measured in degrees, to each nlCWFS channel's electric field prior to propagation to the detector. Five instances of differential tip/tilt are simulated for each instance of atmospheric turbulence.

\paragraph{Higher-Frequency (HF) NCP Aberrations}
\label{sec:HFNCP_methods}


Imperfect optical surfaces (static) and local temperature fluctuations (dynamic) can create higher-frequency (HF) NCP aberrations. We simulate HF NCP aberrations by applying a unique tip-and-tilt-removed phase screen to each nlCWFS channel electric field prior to propagation to the detector. The phase screen RMS is held constant across the channels to normalize HF NCP aberration magnitude. The HF NCP phase screens follow a power spectral density (PSD) of the form
\begin{equation}
    \textrm{PSD} = \frac{1}{1 + (k/k_0)^n},
\end{equation}

\vspace{0.1cm}

\noindent where $k$ is spatial frequency (cycles/m), $n$ is the PSD index and is set to 3, and $k_0$, which helps define the prevalence of high spatial frequencies in the PSD, is set to 10 cycles/m to represent a fold mirror \cite{Shaklan:06}. Five instances of HF NCP aberrations are simulated for each instance of atmospheric turbulence. We vary the HF NCP phase screen RMS to simulate different magnitudes of HF NCP aberrations.

\subsubsection{Servo Lag}\label{sec:SL_methods}

Atmospheric turbulence continues to evolve during servo lag, such that even perfect spatial reconstructions will produce wavefront residuals when corrected with a deformable mirror (DM) that experiences measurable delay. When combined with high wind speeds or rapidly evolving conditions, sufficiently severe servo lag can prohibit AO correction altogether. Servo lag is of particular importance to the nlCWFS given its nonlinear (iterative) reconstruction methods. 

We define servo lag as the time elapsed between read-out at the WFS detector and application of the wavefront correction to the DM. For simplicity, we assume servo lag to be composed entirely of WFS reconstruction runtime. We quantify the RMS WFE resulting from different values of servo lag using the frozen flow turbulence model \cite{Poyneer:09}. We simulate frozen flow turbulence by translating a static phase aberration across the telescope primary at a constant wind speed of 12.5 m/s \cite{Mateen:15}. 

\section{Results}\label{sec:results}

We use RMS WFE to quantify inaccuracies in wavefront reconstruction by calculating the difference between a reconstruction and the known input wavefront. The standard deviation of the RMS WFE measurements across the thirty instances of atmospheric turbulence is used as the statistical uncertainty. To maintain fair and consistent comparisons between sensors and spatial samplings, two adjustments are made to the reconstructions prior to calculating RMS WFE: (i) we interpolate all reconstructions, both the 32$\times$32 and 64$\times$64 classes, to the same 64$\times$64 sample grid; and (ii) set the outer $\sim$6$\%$ of the reconstruction pupil, in units of the pupil radius, to zero to avoid including high-variance edge effects. We separately quantify WFE caused by environmental and systematic effects ($\S$\ref{sec:PN_methods}, $\S$\ref{sec:FBDD_methods} - $\S$\ref{sec:SL_methods}) from WFE caused by inaccuracies in the sensors' reconstruction algorithms ($\S$\ref{sec:AE_methods}). We assume all error sources to be random and independent, justifying the use of quadrature addition for constructing an error budget \cite{Gilles:08} ($\S$\ref{sec:budget}).

\subsection{Algorithmic Errors}\label{sec:AE_results}

\begin{figure}
    \centering
    \includegraphics[width=\textwidth]{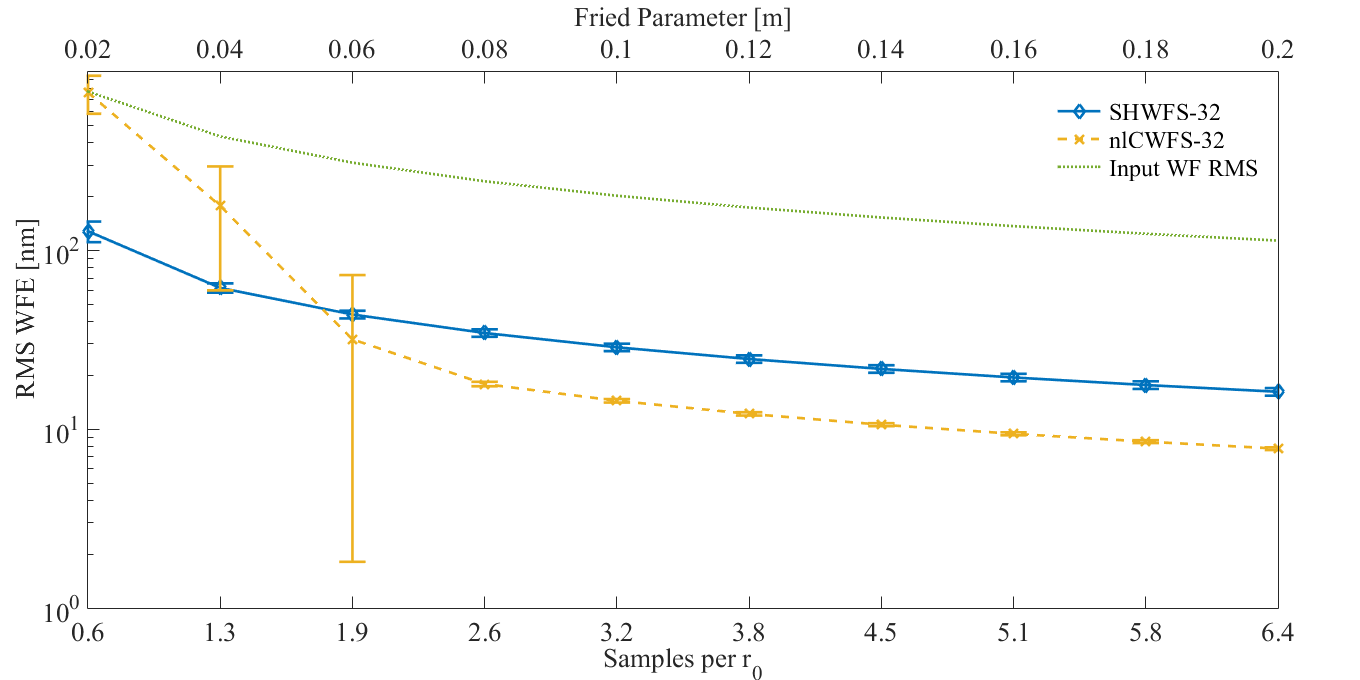}
    \caption{Comparison of algorithmic errors with the SHWFS-32 and nlCWFS-32 in the presence of Kolmogorov turbulence using otherwise noiseless sensors.}
    \label{fig:sense_errors32}
\end{figure}

\begin{figure}
    \centering
    \includegraphics[width=\textwidth]{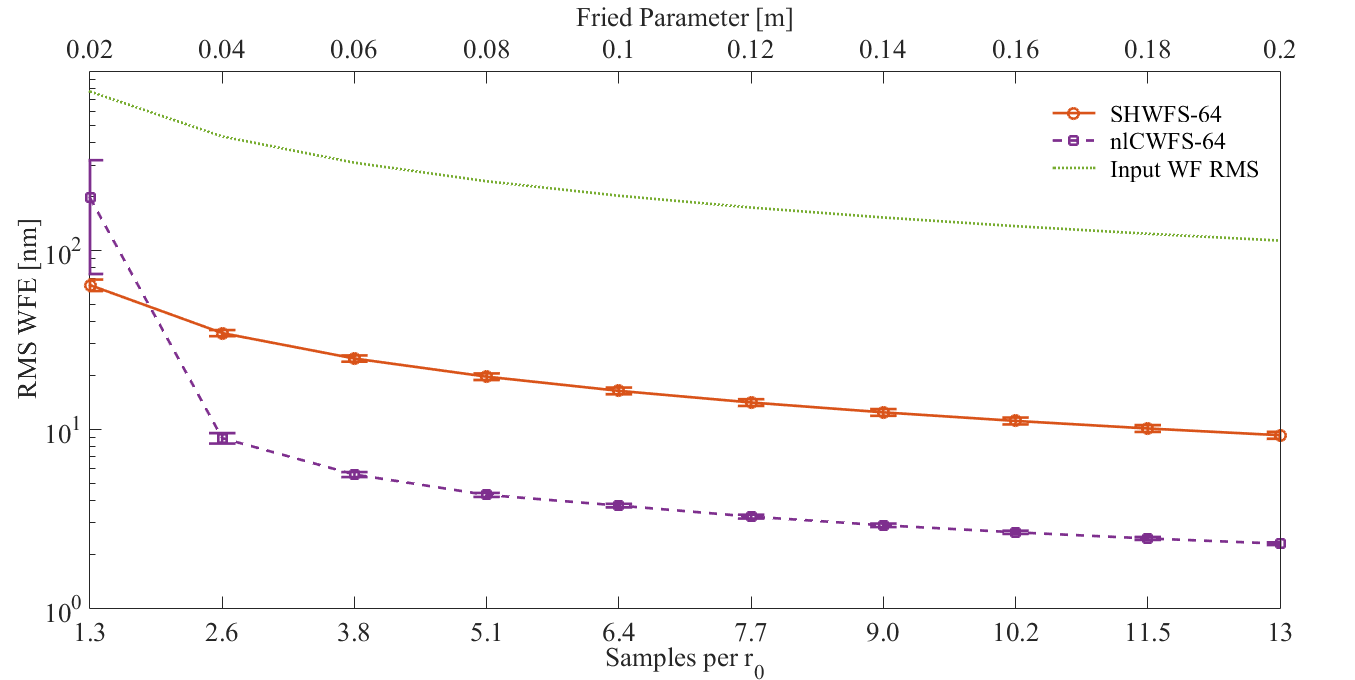}
    \caption{Comparison of algorithmic errors with the SHWFS-64 and nlCWFS-64 in the presence of Kolmogorov turbulence using otherwise noiseless sensors.}
    \label{fig:sense_errors64}
\end{figure}

\begin{figure}
    \centering
    \includegraphics[width=\textwidth]{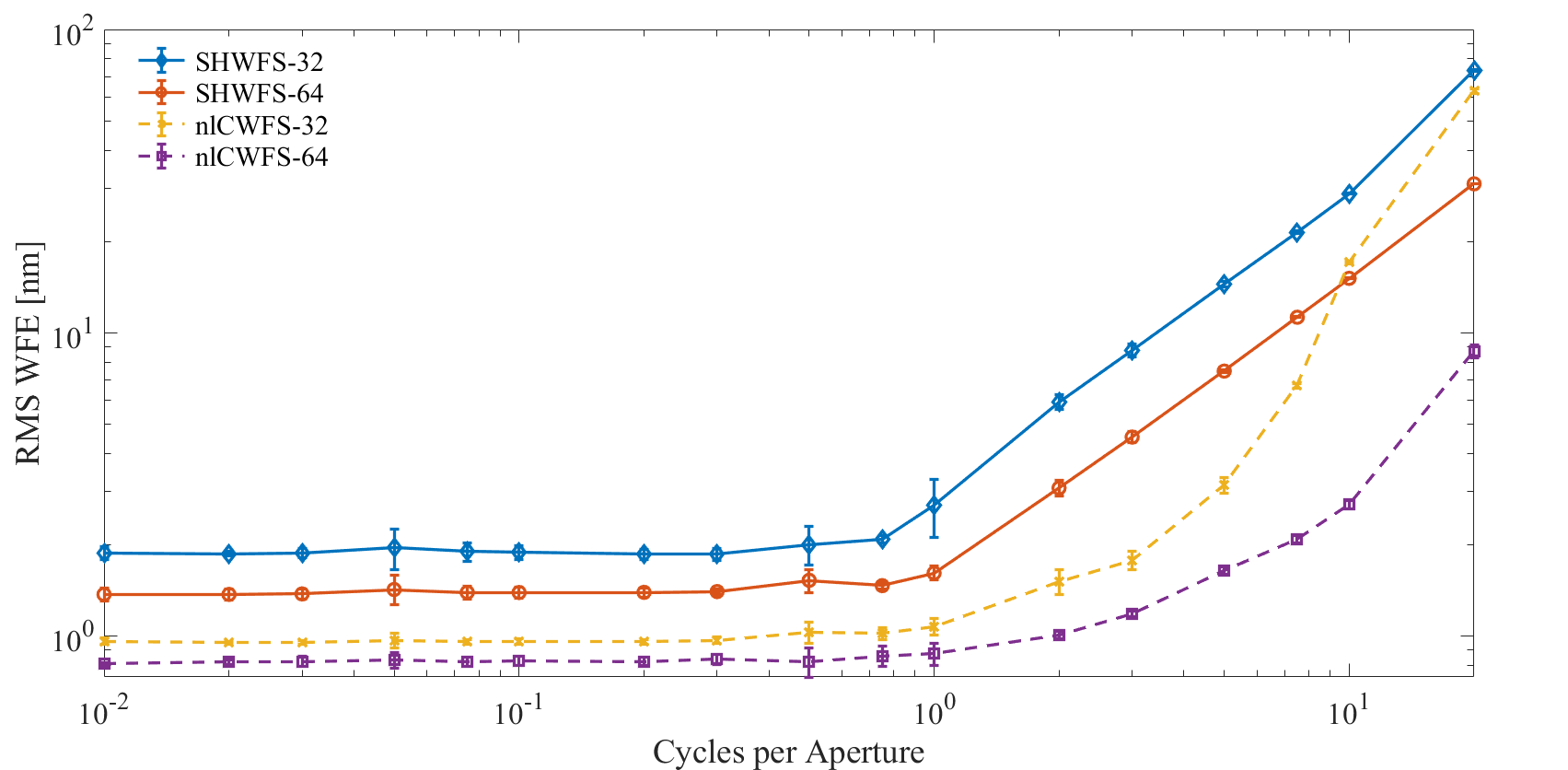}
    \caption{Comparison of algorithmic errors with the SHWFS and nlCWFS as a function of phase distortion spatial frequency.}
    \label{fig:spatfreq}
\end{figure}

Figures \ref{fig:sense_errors32} and \ref{fig:sense_errors64} compare nlCWFS and SHWFS algorithmic errors in the presence of Kolmogorov turbulence. The Fried parameter ranges from $r_0=2$ cm to $r_0=20$ cm, corresponding to a spatial sampling of 0.6 - 6.4 samples per $r_0$ for the nlCWFS-32 and 1.3 - 13 samples per $r_0$ for the nlCWFS-64. No other error sources (photon noise, read noise, etc.) were simulated for Figs. \ref{fig:sense_errors32} - \ref{fig:spatfreq}.

We find that when provided a sufficient number of spatial samples per $r_0$, the nlCWFS-32 and nlCWFS-64 consistently produce more accurate reconstructions than the SHWFS-32 and SHWFS-64, respectively. However, when spatial samples fall below $\approx$2.5 samples per $r_0$, the nlCWFS has difficulty reconstructing the electric field and produces large algorithmic errors. A similar result has been found by Letchev et al., 2022 \cite{Letchev:22}. When provided 2.6 or more samples per $r_0$, the nlCWFS-32 outperforms the SHWFS-32 by a factor of $\approx$2.0$\times$ and the nlCWFS-64 outperforms the SHWFS-64 by a factor of $\approx$4.2$\times$.

Figure \ref{fig:spatfreq} compares nlCWFS and SHWFS algorithmic errors in the presence of sinusoidal phase distortions with peak-to-valley measurements of $\lambda/3$. We vary the spatial frequency of the phase distortion across 4 orders of magnitude. We find that the nlCWFS, when compared to the SHWFS, offers more accurate reconstructions of all spatial frequencies. At low frequencies ($<1$ cycle per aperture), the nlCWFS-32 is more accurate than the SHWFS-32 by a factor of $\approx$2.0$\times$, and the nlCWFS-64 is more accurate than the SHWFS-64 by a factor of $\approx$1.7$\times$. At higher frequencies, the nlCWFS-32 is more accurate than the SHWFS-32 by up to a factor of $\approx$4.9$\times$, and the nlCWFS-64 is more accurate than the SHWFS-64 by up to a factor of $\approx$5.6$\times$. The nlCWFS's ability to produce more accurate reconstructions follows from the sensor's design: the nlCWFS's utilization of full-beam interference enables the sensor to collect information on a broad range of spatial frequencies. However, we find that the nlCWFS-32 experiences a rapid increase in RMS WFE with increasing spatial frequency. This loss in reconstruction accuracy is a function of the inner measurement planes' \emph{z}-locations, which govern the sensor's ability to capture high spatial frequency content \cite{Letchev:22}.

The SHWFS-64 and nlCWFS-64 produce the smallest algorithmic errors of their respective designs in Figs. \ref{fig:sense_errors32} - \ref{fig:spatfreq}. As with any sensor, increasing spatial sampling across the beam improves reconstruction accuracy, establishing a fundamental performance limit in the absence of other noise sources. However, given a decreasing atmospheric power spectrum, one would expect the benefits of increasing the WFS sampling to reach a point of diminishing returns. Eventually, further increases in sampling would enable the capture of high spatial frequencies that contribute only negligibly to overall WFE. Thus, while increasing the SHWFS and nlCWFS sampling past 64$\times$64 may further reduce algorithmic errors, the minimal improvement in WFE would be lost in the presence of moderate noise due to the corresponding loss in the SNR. Photon noise, systematic errors, and environmental effects are studied in the following sections.

\subsection{Photon Noise}
\label{sec:PN_results}

\begin{figure}
    \centering
    \includegraphics[width=\textwidth]{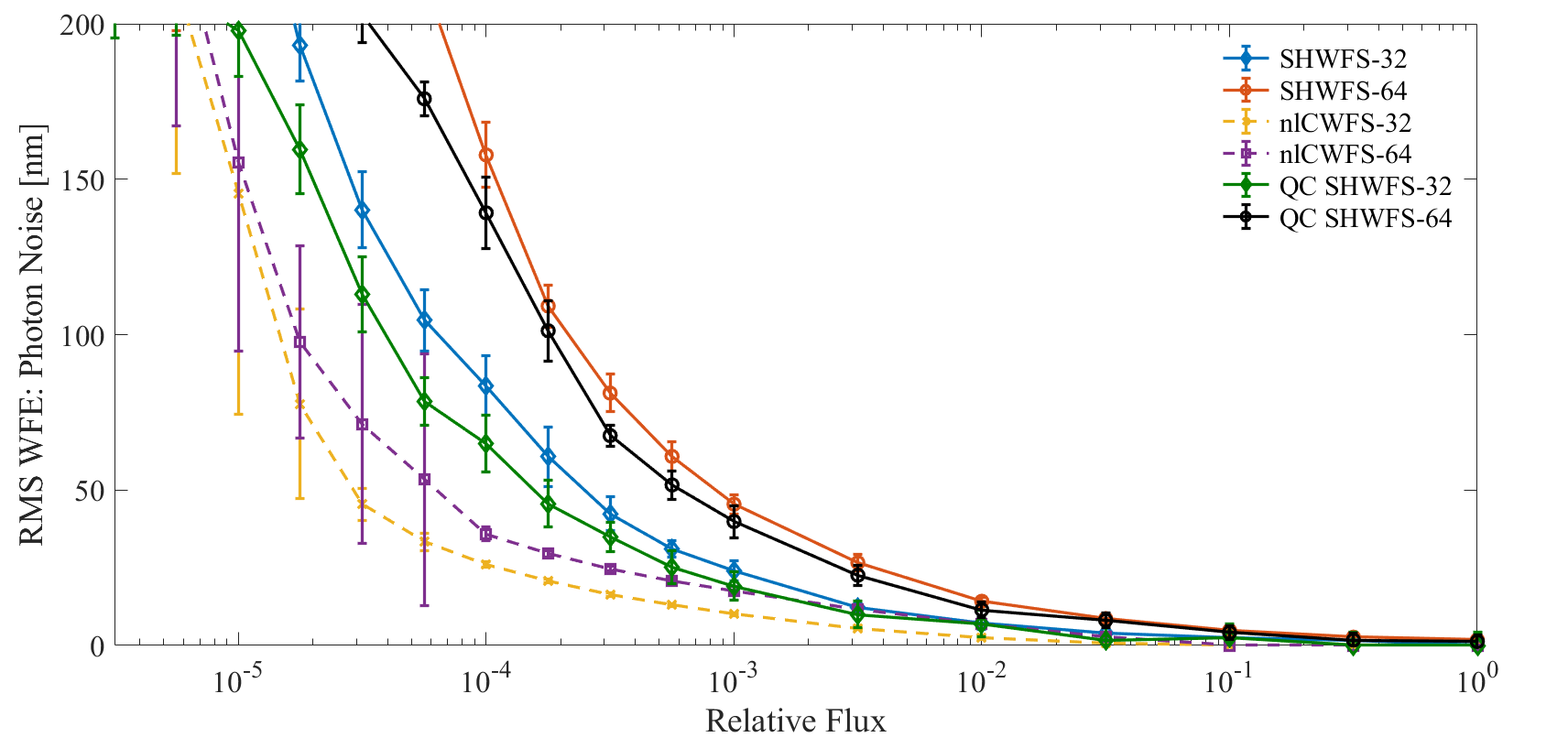}
    \caption{Comparison of the SHWFS and nlCWFS in the presence of photon noise.} 
    \label{fig:pn}
\end{figure}

Figure \ref{fig:pn} compares nlCWFS and SHWFS performance in the presence of photon noise. Diminishing relative flux ($F$) levels are simulated across five orders of magnitude. Of the four sensors studied, we find that the nlCWFS-32 offers the most favorable performance, while the SHWFS-64 offers the least favorable performance. At a relative flux of $F=10^{-4}$, the nlCWFS-32 generates $\approx$3.2$\times$ lower RMS WFE than the SHWFS-32 (26 nm versus 83 nm), and the nlCWFS-64 generates $\approx$4.4$\times$ lower RMS WFE than the SHWFS-64 (36 nm versus 158 nm). Overall, we find that the nlCWFS generates less RMS WFE than the SHWFS for all simulated flux levels. This finding --- that the nlCWFS is more photon-efficient than an equivalent SHWFS --- has been corroborated by a number of numerical studies and laboratory experiments \cite{Guyon:10,Letchev:22,Crass:14,Mateen:15,Crepp:20}; it is the consequence of the nlCWFS using the wave nature of light whereas the SHWFS does not interfere light between subapertures, sacrificing sensitivity to the low spatial frequencies that dominate Kolmogorov turbulence.


We find that the 32$\times$32 sampled sensors generate lower RMS WFE than 64$\times$64 sampled sensors of the same family as a result of the increased SNR. When conserving photons, the flux per sample is $4\times$ higher for the 32$\times$32 case than the 64$\times$64 case. Although 32$\times$32 and 64$\times$64 are common SHWFS configurations, frequently used to help with reconstruction speed because they are multiples of $2^N$, the SHWFS's division of light also depends on spatial sampling in the focal plane. 

As an additional assessment of the efficiency of the sensors studied, we model a SHWFS quad-cell (`QC') configuration that includes only 2$\times$2 pixels behind each lenslet. To ensure accurate results, we establish a precise look-up-table based on centroid calibration data calculated using \emph{WaveProp}'s Shack-Hartmann calibration methods. Otherwise, the same parameters as above were used. Figure \ref{fig:pn} shows that despite the increased SNR of the SHWFS quad-cell configuration (i.e. compared with an equivalent SHWFS that uses more pixels) the nlCWFS still offers better performance for both the 32$\times$32 and 64$\times$64 cases. Thus, reducing the SHWFS's pixel sampling improves sensitivity to faint sources but not beyond the inherent sensitivity of a nlCWFS.


\subsection{Finite Bit Depth}\label{sec:FBDD_results}

\begin{figure}
    \centering
    \includegraphics[width=\textwidth]{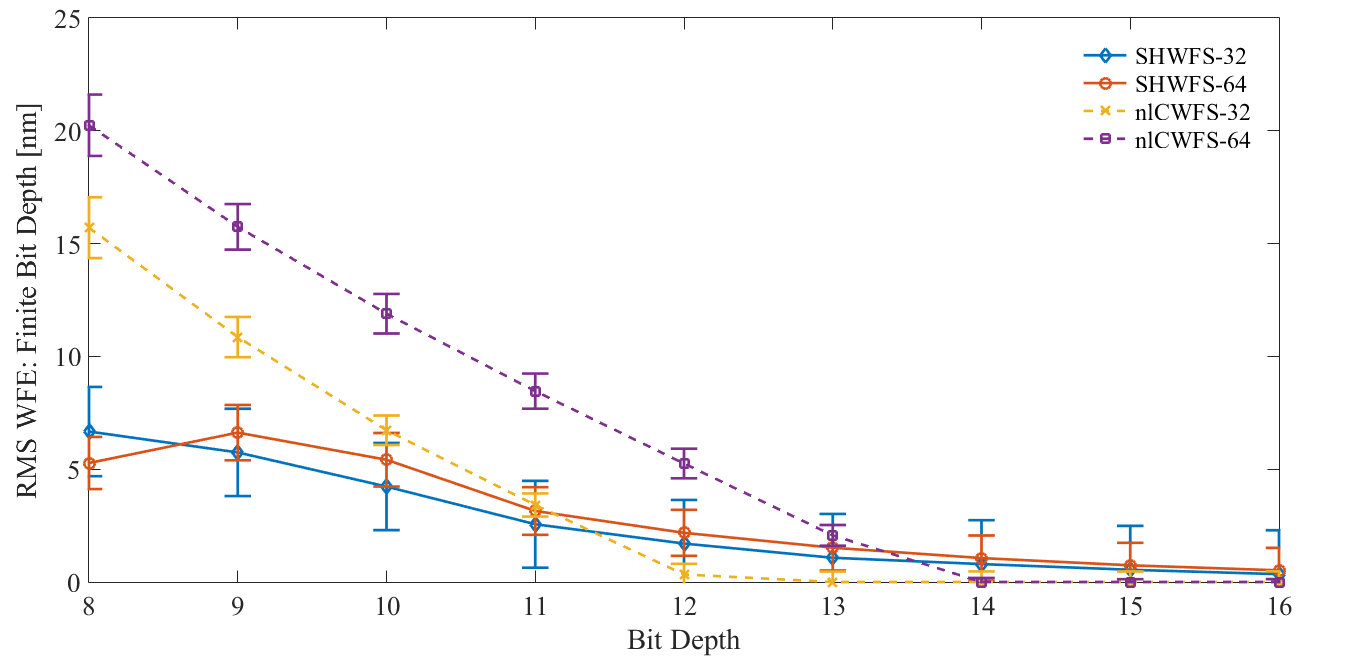}
    \caption{Comparison of the SHWFS and nlCWFS using intensity measurements with finite bit depth.}
    \label{fig:bd}
\end{figure}

Figure \ref{fig:bd} compares the nlCWFS and SHWFS when the sensors are experiencing intensity digitization effects that are limited in precision by finite bit depth ($b$). We find nlCWFS and SHWFS performance to be only minimally impacted by larger bit depths ($b\ge12$), with all sensors incurring less than 6 nm of RMS WFE. However, we find that the SHWFS is less susceptible than the nlCWFS to smaller bit depths ($b\le11$). The nlCWFS, particularly the nlCWFS-64, incurs noticeable WFE at small bit depths: at a bit depth of $b=8$, the nlCWFS-64 incurs 20 nm. The SHWFS experiences minimal WFE across all simulated bit depths, generating only 7 nm at a bit depth of $b=8$. We conclude that a detector with large dynamic range is more important for the nlCWFS than the SHWFS. Bit depths smaller than $b\approx11$ are expected to contribute appreciably to the WFE budget of the nlCWFS. Meanwhile, applications involving the SHWFS can assign less prioritization on detector bit depth.

A potentially important caveat is that increasing the detector bit depth can also reduce image read-out speed, adding to latency. The nlCWFS's nonlinear reconstructor is slower than linear sensors due to its iterative nature. Thus, an optimal detector bit depth should be calculated for applications of the nlCWFS such that WFE from finite bit depth and servo lag are balanced. 

\subsection{Detector Read Noise}
\label{sec:RNnoPED_results}

\begin{figure}
    \centering
    \includegraphics[width=\textwidth]{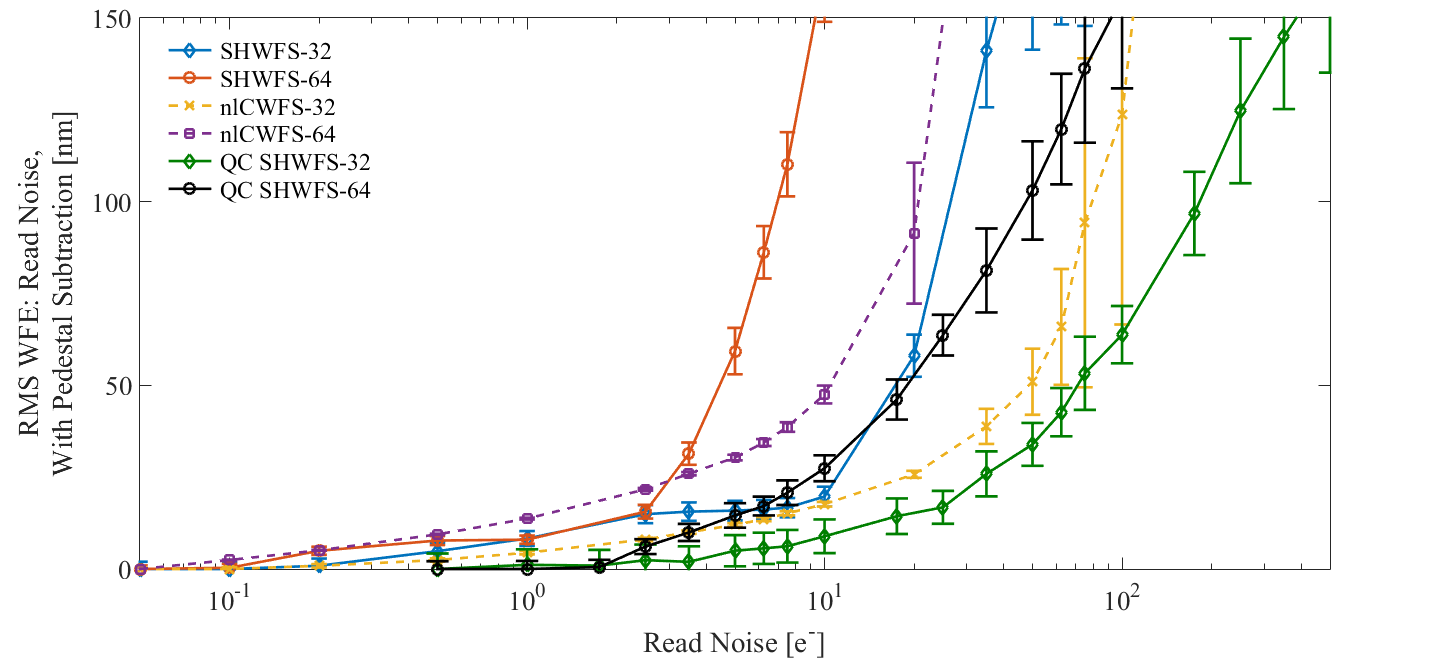}
    \caption{Comparison of the SHWFS and nlCWFS in the presence of detector read noise with optimal pedestal subtraction.}
    \label{fig:mitrn}
\end{figure}

Figure \ref{fig:mitrn} compares the nlCWFS and SHWFS in the presence of detector read noise with optimal pedestal subtraction. Increasing read noise medians are simulated across four orders of magnitude. A flux of $F=10^{-3}$, as shown in Fig. \ref{fig:pn}, is used for each read noise simulation. Of the four sensors studied, we find that the nlCWFS-32 offers the most favorable performance and the SHWFS-64 offers the least favorable performance. At a read noise median of 5$e^-$, the nlCWFS-32 generates $\approx$1.3$\times$ lower RMS WFE than the SHWFS-32 (12 nm versus 16 nm), and the nlCWFS-64 generates $\approx$2.0$\times$ lower RMS WFE than the SHWFS-64 (30 nm versus 59 nm). For read noise medians below 2.5$e^-$, we find that all WFS configurations offer a similar low-WFE performance ($< 22$ nm RMS). We conclude that a detector with low read noise is more important for the SHWFS than the nlCWFS when optimal pedestal subtraction methods are employed. While Crepp et al., 2020\cite{Crepp:20} previously found using lab experiments that the nlCWFS is more susceptible to read noise than a comparable SHWFS, this analysis did not include optimized pedestal subtraction.

We find that the 32$\times$32 sampled sensors generate lower RMS WFE than 64$\times$64 sampled sensors of the same family as a result of the increased measurement SNR. However, the SNR of a SHWFS measurement also depends on the sensor's spatial sampling in the focal plane. To further assess SHWFS performance in the presence of read noise with optimal pedestal subtraction, we once again model a SHWFS QC configuration with 2$\times$2 pixels behind each lenslet. 

Figure \ref{fig:mitrn} shows that the increased SNR of the SHWFS quad-cell configuration improves the sensor's resilience to read noise with optimal pedestal subtraction beyond that of an equivalent nlCWFS. At a read noise median of 5$e^-$, the QC SHWFS produces $\approx$2.0$\times$ less RMS WFE than the nlCWFS in the 64$\times$64 case (15 nm versus 30 nm) and $\approx$2.4$\times$ less in the 32$\times$32 case (5 nm versus 12 nm). However, decreasing the SHWFS focal plane sampling reduces the sensor's reconstruction accuracy, introducing additional algorithmic errors. For example, reducing the number of pixels behind each SHWFS lenslet from 8$\times$8 pixels to 2$\times$2 pixels increases the sensor's algorithmic errors by a factor of $\approx$4.6$\times$ at $r_0=2$ cm. Thus, an optimal SHWFS focal plane sampling should be calculated for applications of the SHWFS such that WFE from read noise and algorithmic errors are balanced.



\subsection{Detector Vibrations}\label{sec:VIB_results}

\begin{figure}
    \centering
    \includegraphics[width=\textwidth]{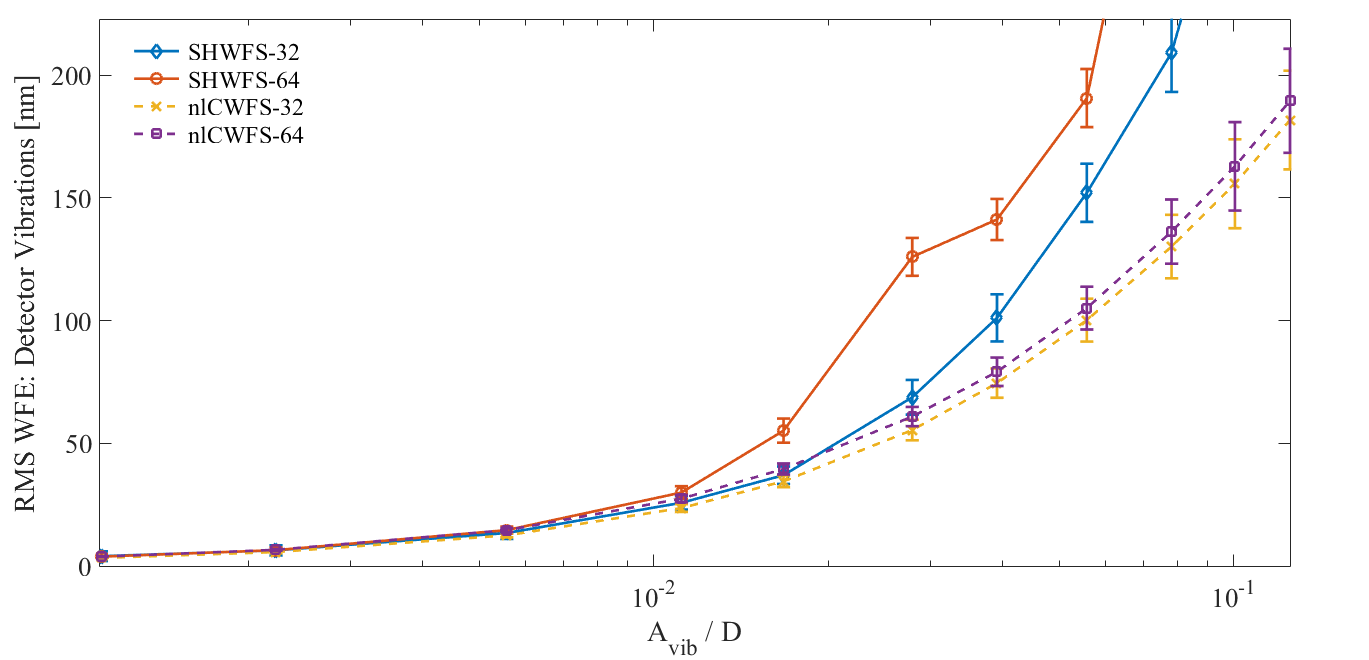}
    \caption{Comparison of the SHWFS and nlCWFS in the presence of detector vibrations. $\textrm{A}_\textrm{vib}$ is the vibration amplitude and $D$ is the WFS beam diameter.}
    \label{fig:vib}
\end{figure}

Figure \ref{fig:vib} compares nlCWFS and SHWFS performance in the presence of detector vibrations. We find that all four sensor configurations are robust to low-amplitude detector vibrations, provided the vibrations are smaller than 1\% of the beam diameter ($D$). Vibration amplitudes larger than 1\% of $D$ create an increasingly challenging reconstruction environment, resulting in RMS WFE values that approach a large fraction of a wave. 

In all cases simulated, we find that the nlCWFS is more robust to vibrations than the SHWFS. At a vibration amplitude equal to 4\% of D, the nlCWFS-32 and nlCWFS-64 outperform the SHWFS-32 and SHWFS-64 by factors of $\approx$1.4$\times$ and $\approx$1.8$\times$, respectively. Vibration sensitivity in the SHWFS can be attributed to its use of a lenslet array. When the vibration amplitude exceeds the lenslet pitch, a displacement of the local wavefront to its neighbor occurs. This problem is particularly troublesome for the SHWFS-64 design, which has lenslets half as large as the SHWFS-32. Meanwhile, the nlCWFS design benefits from using a continuous array of detector pixels, which allows the sensor to have a consistent response to vibrations of increasing amplitude. The above analysis was performed with uncompensated vibrations. Beam tracking technology may be used in practice to reduce the RMS WFE reported in Fig. \ref{fig:vib} for both WFS designs \cite{Spencer:18}.

\subsection{Non-Common-Path (NCP) Aberrations in the nlCWFS}

\subsubsection{Differential Tip/Tilt}
\label{sec:DT_results}

\begin{figure}
    \centering
    \includegraphics[width=\textwidth]{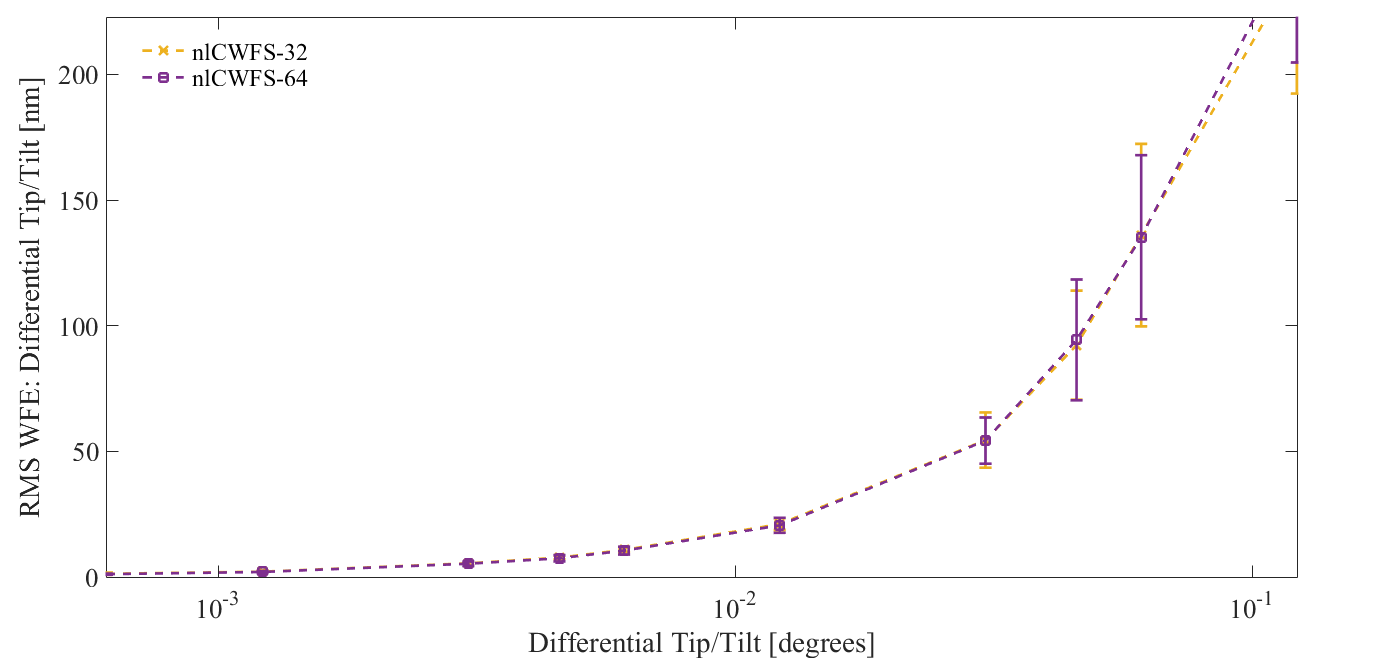}
    \caption{nlCWFS performance in the presence of differential tip/tilt. Differential tip/tilt is applied at the pupil plane prior to propagation to the measurement planes and is measured in degrees.}
    \label{fig:difftilt}
\end{figure}

Figure \ref{fig:difftilt} shows the performance of the nlCWFS in the presence of differential tip/tilt. In terms of absolute measurements, we find that individual nlCWFS beam channels should be directed more precisely than $\approx$0.01$\degree$ (to the nearest order of magnitude) to avoid an appreciable contribution to the sensor's WFE budget. However, such tight tolerances on optical alignment can be relieved by tracking the measurement plane centroids of each beam in the nlCWFS reconstruction software. For example, using a beam diameter of 0.5 mm and propagation distance of 1 cm, a differential tip/tilt of 0.01$\degree$ corresponds to a displacement of the measurement plane centroids by $\sim$0.22 pixels for the nlCWFS-64 and $\sim$0.11 pixels for the nlCWFS-32. Such displacements, while small, can be monitored and adjusted prior to WFS operation. The above analysis considers only static alignment effects. Dynamic effects, such as optics jitter, must be aggressively mitigated or corrected with high frequency centroid measurements to avoid differential tip/tilts greater than 0.01$\degree$ and substantial WFE residuals. 

Tight differential tip/tilt constraints can be made less severe by increasing the distances between the pupil and nlCWFS measurement planes. Lengthening the measurement plane distances increases the number of pixels behind each measurement plane image, which enables more precise centroid adjustments. However, any reductions in RMS WFE gained by loosening the differential tip/tilt tolerances need to be balanced with the RMS WFE incurred due to the decrease in the SNR that increased measurement plane samplings produce in the presence of noise.

\subsubsection{Higher-Frequency NCP Aberrations}
\label{sec:HFNCP_results}

\begin{figure}
    \centering
    \includegraphics[width=\textwidth]{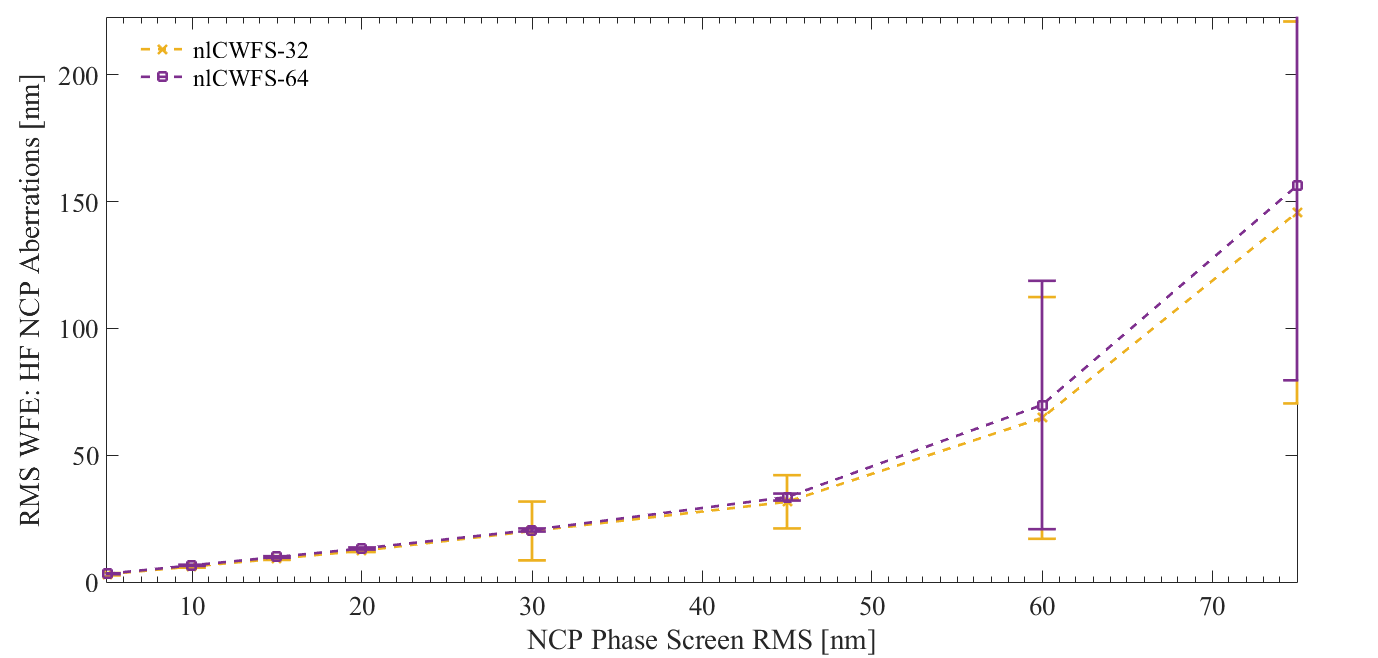}
    \caption{nlCWFS performance in the presence of HF NCP aberrations.}
    \label{fig:inco}
\end{figure}

Figure \ref{fig:inco} shows the performance of the nlCWFS in the presence of HF NCP aberrations. We find that HF NCP aberrations must be limited to an effective RMS of 20 nm to avoid an appreciable contribution to the nlCWFS's WFE budget. Larger HF NCP aberrations introduce significant WFE, with the nlCWFS incurring increasingly large amounts of RMS WFE for effective HF NCP RMS values greater than 45 nm. Thus, applications of the nlCWFS must minimize aberrations produced by non-common optics by (i) decreasing the number of optics in each nlCWFS channel and (ii) using optics with pristine surface flatness. Temperature uniformity must also be maintained across the nlCWFS channels to prevent additional HF NCP aberrations.

\subsection{Servo Lag}
\label{sec:SL_results}

\begin{figure}
    \centering
    \includegraphics[width=\textwidth]{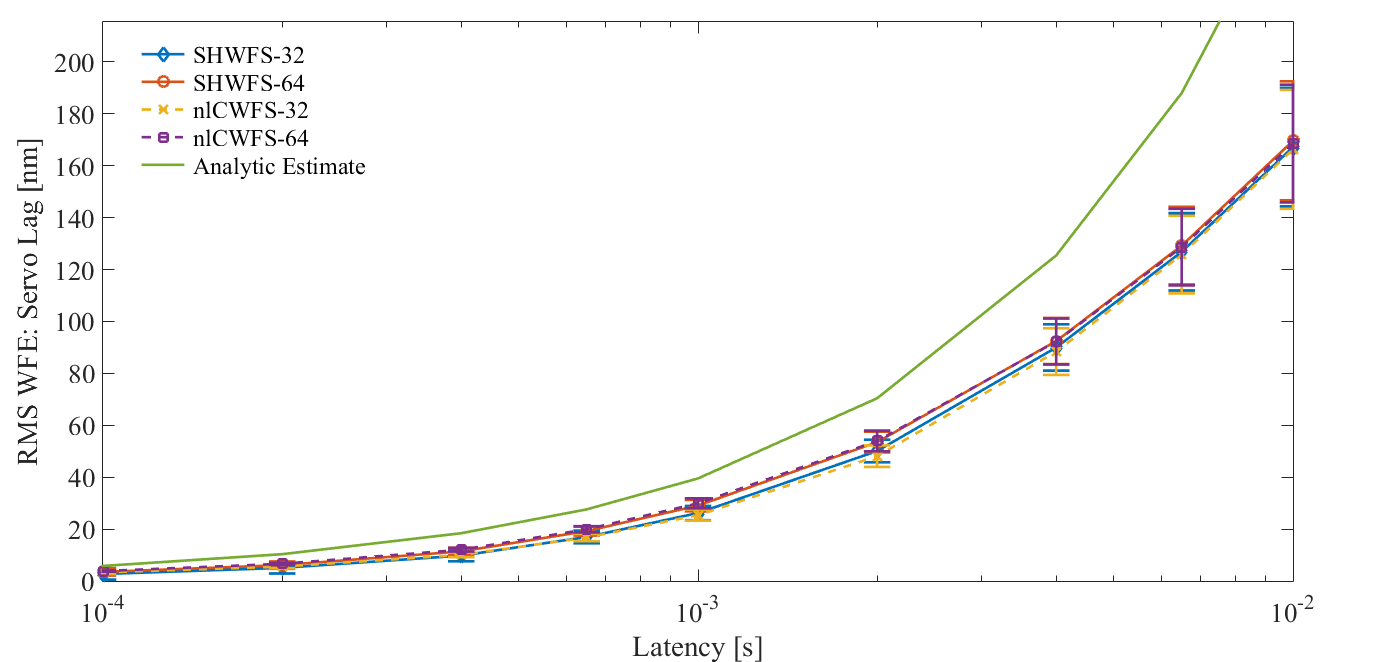}
    \caption{Comparison of SHWFS and nlCWFS servo lag WFE in the presence wind with speed of $V_{\textrm{wind}}=12.5$ m/s.}
    \label{fig:latency}
\end{figure}

Figure \ref{fig:latency} compares the nlCWFS and SHWFS in the presence of servo lag and wind. We assume a constant wind speed of 12.5 m/s. Section \ref{sec:benchmarks} provides representative timescales for SHWFS and nlCWFS servo lag. Assuming complete independence from algorithmic errors, the simulated servo lag WFE should depend only on WFS latency and wind speed, not the WFS design. Consistent with this hypothesis, we find that each WFS configuration follows a similar RMS WFE curve. The small differences between the WFS configurations arise from an imperfect independence between the servo lag WFE and the algorthmic error WFE. However, the differences are within the range of uncertainty and can be ignored. 

The results presented in Fig. \ref{fig:latency} do not account for the differences in operating frequency between the WFSs. In practice, the nlCWFS reconstructor is slower than the SHWFS's due to its iterative, nonlinear methods. Thus, the nlCWFS is anticipated to produce larger reconstruction runtimes and servo lag RMS WFE than the SHWFS, particularly the higher-sampled nlCWFS-64. In Sec. \ref{sec:budget}, we use scaling arguments to estimate timing delays incurred with each sensor.

For reference, Fig. \ref{fig:latency} includes an analytic estimate of the servo lag RMS WFE. To calculate servo lag RMS WFE ($\sigma_{SL}$) as a function of WFS latency and wind speed, we combine the following equations \cite{Fried:90,Tyler:94},

\begin{equation}
\sigma_{SL}^2 = 28.44(f_{G}\Delta t)^{5/3}
\end{equation}
\begin{equation}
f_G = 0.43\frac{V_{\textrm{wind}}}{r_0},
\end{equation}

\noindent where $\sigma_{SL}$ is measured in radians, $f_G$ is the Greenwood frequency in Hz, $\Delta t$ is the latency in seconds, $r_0$ is the Fried parameter in meters, and $V_{\textrm{wind}}$ is the average wind speed in m/s. The derivation of the analytic estimate \cite{Fried:90} and the frozen-flow simulations used to produce the data in Fig. \ref{fig:latency} differ in their treatment of tip and tilt, which dominate the atmosphere's power spectrum. The analytic estimate derivation includes all tip and tilt while the simulations use solely tip-and-tilt-removed wavefronts. Thus, the simulations should report less servo lag RMS WFE than the analytic estimate. Consistent with this hypothesis, we find that the servo lag trendlines of Fig. \ref{fig:latency} closely match the shape of the analytic estimate trendline while also reporting consistently lower RMS WFE values.

\section{Wavefront Error Budget}\label{sec:budget}

We combine the results of the previous sections into a WFE budget for each sensor and spatial sampling configuration (Tables \ref{tab:WFE_budget_32} and \ref{tab:WFE_budget_64}). Additionally, we include estimates of the anticipated RMS WFE for three error sources not simulated in this study: DM fitting error, SHWFS spot size calibration errors, and SHWFS uncalibrated lenslet array (ULA) errors.

For DM fitting errors, we calculate the anticipated RMS WFE using the equations provided by Tyson and Frazier, 2022 \cite{Tyson:22}. The RMS WFE is dependent on the spacing between DM actuators, $d_{s}$, and $r_0$ as follows:

\begin{equation}
    \sigma^{2}_{\textrm{fit}} = \mu\left( \frac{d_{s}}{r_0} \right)^{5/3},
\end{equation}

\noindent where $\mu$ is a constant that depends on the DM response function \cite{Tyson:22}. We assume a 32$\times$32 WFS is combined with a DM with 32$\times$32 actuators across the pupil and a 64$\times$64 WFS is combined with a DM with 64$\times$64 actuators across the pupil. 

\begin{table*}[t]
\centering
    \def\arraystretch{1.4}
    \caption{nlCWFS-32 and SHWFS-32 RMS WFE budget.}
    \begin{tabular}{lcc|cc}
    \hline
    \hline
    ~ & \multicolumn{2}{c}{Optimistic [nm]} & \multicolumn{2}{c}{Pessimistic [nm]}      \\
    \hline
    \hline
     Error Source                 & nlCWFS         & SHWFS        & nlCWFS        & SHWFS   \\
     \hline
     Algorithmic Errors           &  8           &  16          & 18          &  35     \\ 
     Photon Noise                 &  13          &  31          & 26          &  84     \\
     Finite Bit Depth             &  0           &  1           & 0           &  2      \\ 
     Read Noise                   &  4           &  8           & 18          &  20     \\ 
     Detector Vibrations          &  11          &  12          & 31          &  33     \\ 
     DM Fitting Errors            &  15          &  15          & 19          &  19     \\
     Differential Tip/Tilt        &  2           &  --          & 31          &  --     \\ 
     HF NCP Aberrations           &  16          &  --          & 43          &  --     \\
     Spot Size Calibration Errors &  --          &  10          & --          &  40     \\
     Uncalibrated LA Errors       &  --          &  10          & --          &  30     \\
    \hline
    \hline
    Quadrature Sum (Spatial) & 29          & 43           & 74          & 112           \\
    Strehl Ratio (Spatial)   & 0.89        & 0.77         & 0.47        & 0.17          \\
    \hline
    \hline
    Servo lag                & 24          & 7            & 35          & 8             \\ 
    \hline
    \hline
    Quadrature Sum (Total)   & 38          & 44           & 82          & 113           \\
    Strehl Ratio (Total)     & 0.82        & 0.76         & 0.39        & 0.17          \\    
    \hline
    \hline
    \end{tabular}
    \label{tab:WFE_budget_32}
\end{table*}

\begin{table*}[t]
\centering
    \def\arraystretch{1.4}
    \caption{nlCWFS-64 and SHWFS-64 RMS WFE budget.}
    \begin{tabular}{lcc|cc}
    \hline
    \hline
    ~ & \multicolumn{2}{c}{Optimistic [nm]} & \multicolumn{2}{c}{Pessimistic [nm]}      \\
    \hline
    \hline
     Error Source                 & nlCWFS         & SHWFS        & nlCWFS        & SHWFS   \\
     \hline
     Algorithmic Errors           &  2           &  9           & 4           &  20     \\ 
     Photon Noise                 &  21          &  61          & 36          &  158    \\
     Finite Bit Depth             &  0           &  1           & 5           &  2      \\ 
     Read Noise                   &  14          &  8           & 47          &  163    \\ 
     Detector Vibrations          &  13          &  13          & 36          &  47     \\ 
     DM Fitting Errors            &  9           &  9           & 11          &  11     \\
     Differential Tip/Tilt        &  1           &  --          & 16          &  --     \\ 
     HF NCP Aberrations           &  17          &  --          & 46          &  --     \\ 
     Spot Size Calibration Errors &  --          &  10          & --          &  40     \\
     Uncalibrated LA Errors       &  --          &  10          & --          &  30     \\
    \hline
    \hline
    Quadrature Sum (Spatial) & 34          & 66           & 86          &  238          \\
    Strehl Ratio (Spatial)   & 0.85        & 0.54         & 0.36        &  0.00         \\
    \hline
    \hline
    Servo lag                & 91          & 14           & 120         & 21            \\ 
    \hline
    \hline
    Quadrature Sum (Total)   & 97          & 67           & 148         & 239           \\
    Strehl Ratio (Total)     & 0.27        & 0.53         & 0.05        & 0.00          \\    
    \hline
    \hline
    \end{tabular}
    \label{tab:WFE_budget_64}
\end{table*}


Spot size calibration errors and ULA errors are unique to the SHWFS design. Spot size calibration errors result from poorly calibrated or dynamic spot sizes, which inaccurately inform the SHWFS reconstructor. ULA errors correspond to imperfections in the lenslet array that i) cannot be counteracted through DM calibration and ii) produce light artifacts at the detector that are uncorrelated with the input wavefront. We estimate the anticipated RMS WFE from spot size calibration errors and ULA errors by referring to values provided in a 2008 Keck Adaptive Optics Note \cite{Neyman:08} for the Keck Next Generation AO (NGAO) System and the Thirty Meter Telescope's (TMT's) Narrow Field InfraRed AO System (NFIRAOS).

To consider the range of each error source listed in the WFE budgets and help frame the problem, we establish two representative cases to benchmark each error term, ``optimistic'' and ``pessimistic,'' based on simulations, literature values, and lab experiments. The optimistic and pessimistic cases of each error source are detailed in the following subsection. Evaluated using Figs. \ref{fig:sense_errors32}-\ref{fig:latency}, the resulting RMS WFE values are then added in quadrature to provide estimates of the spatial domain RMS WFE and total (spatial domain plus temporal domain) RMS WFE. We provide estimates of the corresponding Strehl ratios using the Marechal approximation \cite{Ross:09} as an additional measure of WFS performance. Tables \ref{tab:WFE_budget_32} and \ref{tab:WFE_budget_64} summarize the results of the 32$\times$32 and 64$\times$64 WFE budgets, respectively.


\subsection{WFE Benchmarking}\label{sec:benchmarks}

We benchmark the impact of photon noise on WFS performance by selecting representative optimistic and pessimistic relative fluxes of $F=10^{-3.25}$ and $F=10^{-4}$. For all other error sources, we justify the selection of optimistic and pessimistic cases as follows.


For algorithmic errors, we assume that the nlCWFS-32 and nlCWFS-64 offer sufficient pixel sampling ($>$2.5 pixels per $r_0$) to ensure convergence of the wavefront reconstructor. For the design parameters chosen for the study, this range corresponds to $r_0=0.08$ m (pessimistic) and $r_0=0.20$ m (optimistic). Larger $r_0$ values would otherwise correspond to a very quiescent telescope observing site. $r_0=$ 10 cm for the benchmarking of all other error terms.

For finite bit depth, we select representative optimistic and pessimistic cases of 14-bit operation and 12-bit operation. Modern CCD, sCMOS, SWIR, and related cameras and their controllers now tend to offer this level of digitization precision or better.

For detector read noise, we select representative optimistic and pessimistic medians of 1$e^-$ and 10$e^-$. A variety of modern detectors, including those with multi-kHz readout modes, now offer low read noise settings that border on the ability to detect individual photons. Both readout speed and electronic noise will presumably continue to improve.

For detector vibrations, we select representative optimistic and pessimistic vibration amplitudes equal to 0.5\% and 1.5\% of $D$. Further reductions in vibration amplitude are realistic.

For DM fitting errors, we select representative optimistic and pessimistic values for $\mu$ of 0.23 and 0.349, which represent a Gaussian influence function and an influence function that is not constrained at the DM edge, respectively \cite{Tyson:22}.

For differential tip/tilt errors, we assume in the optimistic case that static sources of differential tip/tilt can be calibrated to within 0.01 pixels at the detector with symmetric imaging. Such pristine calibration is possible with turbulence-free measurement plane images and accurate centroiding methods. However, even weak turbulence produces asymmetries in the measurement planes, reducing the precision of the calibration procedure. Further, dynamic sources of differential tip/tilt such as optics jitter require the use of high frequency centroid measurements, which are likely to be less accurate than pre-calibration centroiding methods. Therefore, we assume in the pessimistic case that differential tip/tilt can only be calibrated to within 0.2 pixels at the detector. An optimistic calibration accuracy of 0.01 pixels corresponds to 0.0009$\degree$ of differential tip/tilt for the nlCWFS-32 and 0.00045$\degree$ of differential tip/tilt for the nlCWFS-64; a pessimistic calibration accuracy of 0.2 pixels corresponds to 0.018$\degree$ of differential tip/tilt for the nlCWFS-32 and 0.009$\degree$ of differential tip/tilt for the nlCWFS-64.

For HF NCP aberrations, we use laboratory data to help inform the optimistic and pessimistic cases. Crepp et al., 2020 \cite{Crepp:20} were able to produce SHWFS and nlCWFS wavefront reconstructions that agreed to within $0.09$ waves RMS using commercial off-the-shelf components, which corresponds to 48 nm RMS at a sensing wavelength of 532 nm. Custom optics with smaller surface aberrations can likely be procured. Thus, we select representative optimistic and pessimistic HF NCP phase screen RMS values of 25 nm and 50 nm, respectively.

For spot size calibration errors, we select optimistic and pessimistic RMS WFE values of 10 nm and 40 nm to represent the data reported for the TMT's NFIRAOS and the KECK NGAO System, respectively \cite{Neyman:08}.

For ULA errors, we select a pessimistic RMS WFE value of 30 nm to represent the data reported for the Keck NGAO System \cite{Neyman:08}. We select an optimistic RMS WFE value of 10 nm.


For servo lag, we assume that the nlCWFS (nonlinear) reconstructor is slower than a comparable SHWFS (linear) reconstructor due to the nlCWFS's increased computational complexity. In absolute terms, we assume that the SHWFS-32 has a reconstructor runtime of 200 $\mu$s. We then scale the SHWFS-32 reconstructor runtime to calculate the corresponding SHWFS-64, nlCWFS-32, and nlCWFS-64 reconstructor runtimes. The timing data used to compare reconstruction runtimes includes SHWFS timing data provided by Thorlabs \cite{Thorlabs:NA} and laboratory timing data acquired from previous nlCWFS hardware experiments. We find that, assuming no further optimization, the SHWFS-64 is $\approx$2.5$\times$ slower than the SHWFS-32; the nlCWFS-32 is $\approx$4.3$\times$ slower than the SHWFS-32; and the nlCWFS-64 is $\approx$20$\times$ slower than the SHWFS-32. Using Fig. \ref{fig:latency}, we find that the SHWFS-32's reconstructor runtime of 200 $\mu$s corresponds to 7 nm of servo lag RMS WFE. Similar evaluations of Fig. \ref{fig:latency} with the SHWFS-64, nlCWFS-32, and nlCWFS-64 reconstructor runtimes produce RMS WFE values of 14 nm (500 $\mu$s), 24 nm (868 $\mu$s), and 91 nm (4013 $\mu$s), respectively. We assume these reconstructor runtimes in the optimistic case and estimate the pessimistic case runtimes to be 1.5$\times$ larger.


\subsection{Total WFE Estimates}\label{sec:totalWFEestimates}

The results presented in Table \ref{tab:WFE_budget_32} and Table \ref{tab:WFE_budget_64} show that the nlCWFS generates less WFE than the SHWFS for nearly all simulated spatial domain error sources. Notably, the nlCWFS maintains a distinct advantage in photon noise RMS WFE, incurring between 18 nm and 120 nm less RMS WFE than the SHWFS across the simulated cases. Such improvements in performance are critical for applications involving low illumination, including observations using faint guide stars. However, the SHWFS generates a similar or lesser amount of RMS WFE than an equivalent nlCWFS when both spatial and temporal (servo lag) effects are considered, suggesting improvements to the nlCWFS's operating frequency must be made in order to exploit the sensor's advantages in the spatial domain. For both WFS designs, sensor-specific error sources such as HF NCP aberrations and spot size calibration errors can contribute appreciably to a sensor's error budget if not mitigated to the optimistic case. 



\begin{figure}
    \centering
    \includegraphics[width=\textwidth]{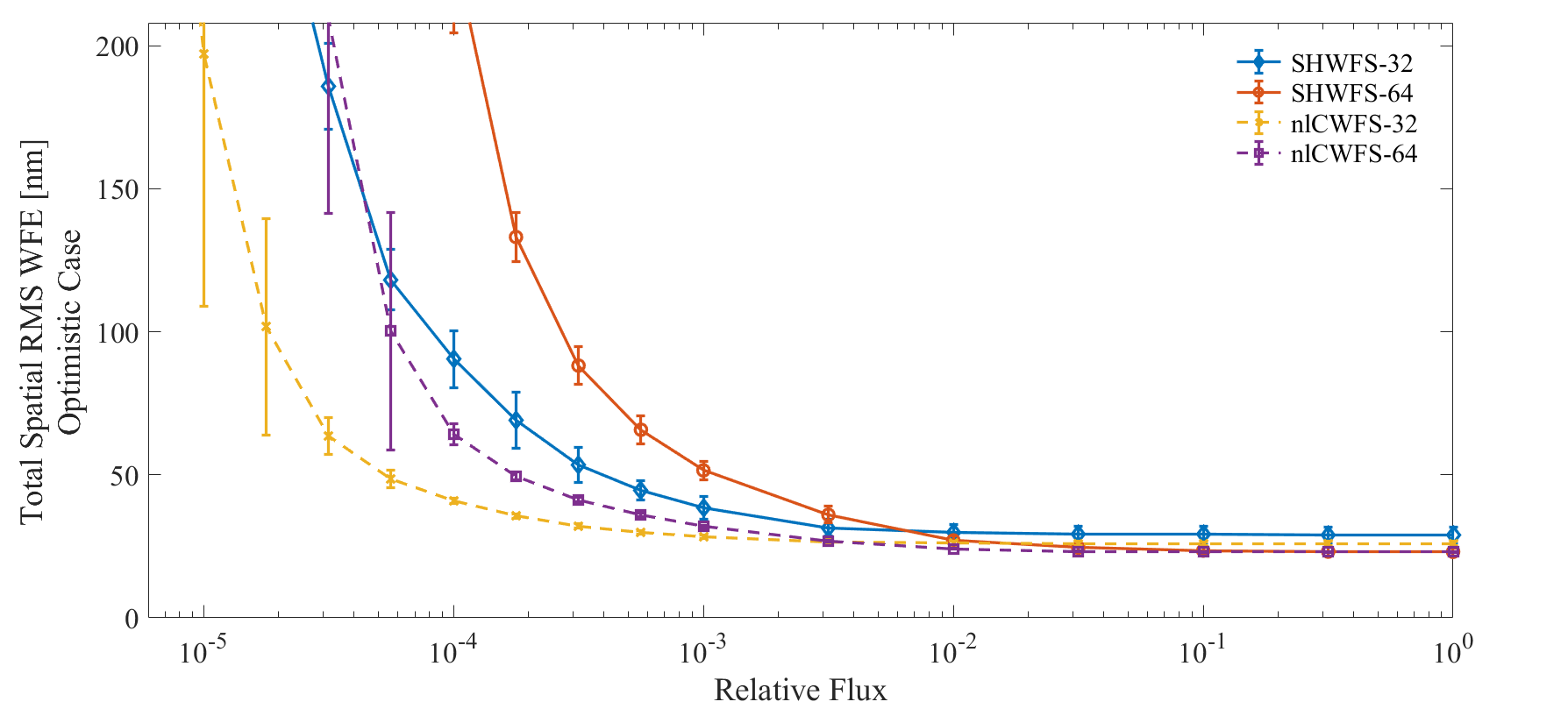}
    \caption{Comparison of the optimistic SHWFS and nlCWFS spatial RMS WFE in the presence of diminishing flux.}
    \label{fig:totalWFEestimates_opt}
\end{figure}

\begin{figure}
    \centering
    \includegraphics[width=\textwidth]{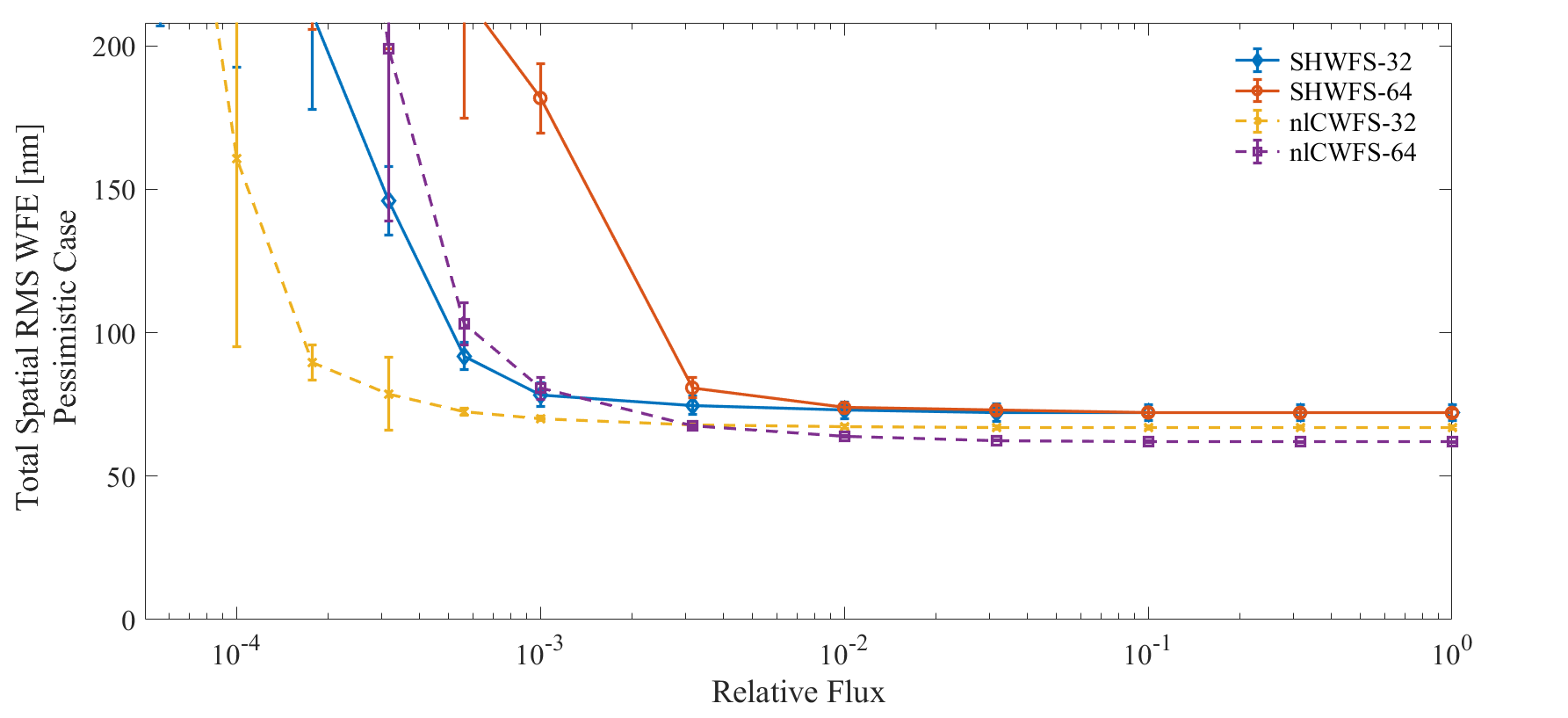}
    \caption{Comparison of the pessimistic SHWFS and nlCWFS spatial RMS WFE in the presence of diminishing flux.}
    \label{fig:totalWFEestimates_pess}
\end{figure}

\begin{figure}
    \centering
    \includegraphics[width=\textwidth]{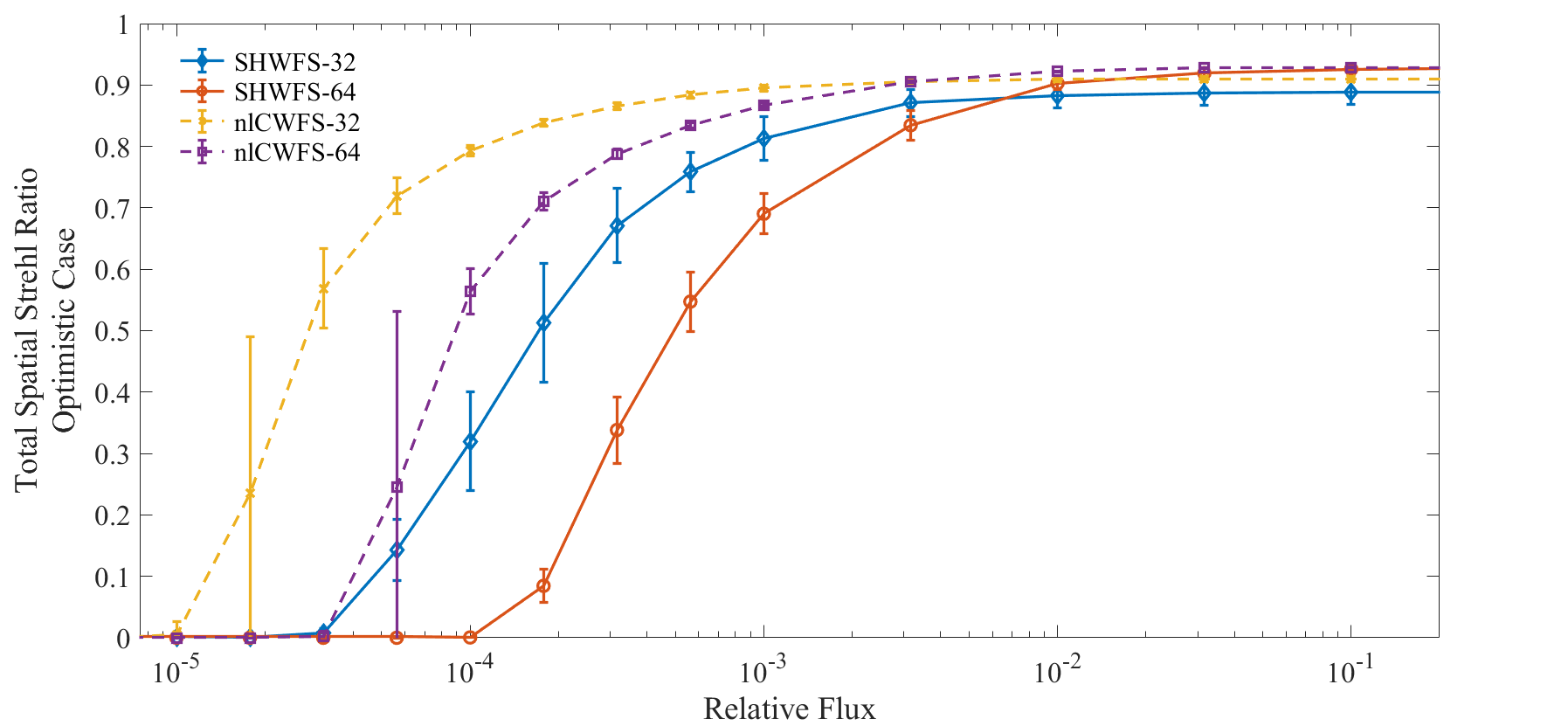}
    \caption{Comparison of the optimistic SHWFS and nlCWFS spatial Strehl ratio in the presence of diminishing flux.}
    \label{fig:totalWFEestimates_opt_strehl}
\end{figure}

\begin{figure}
    \centering
    \includegraphics[width=\textwidth]{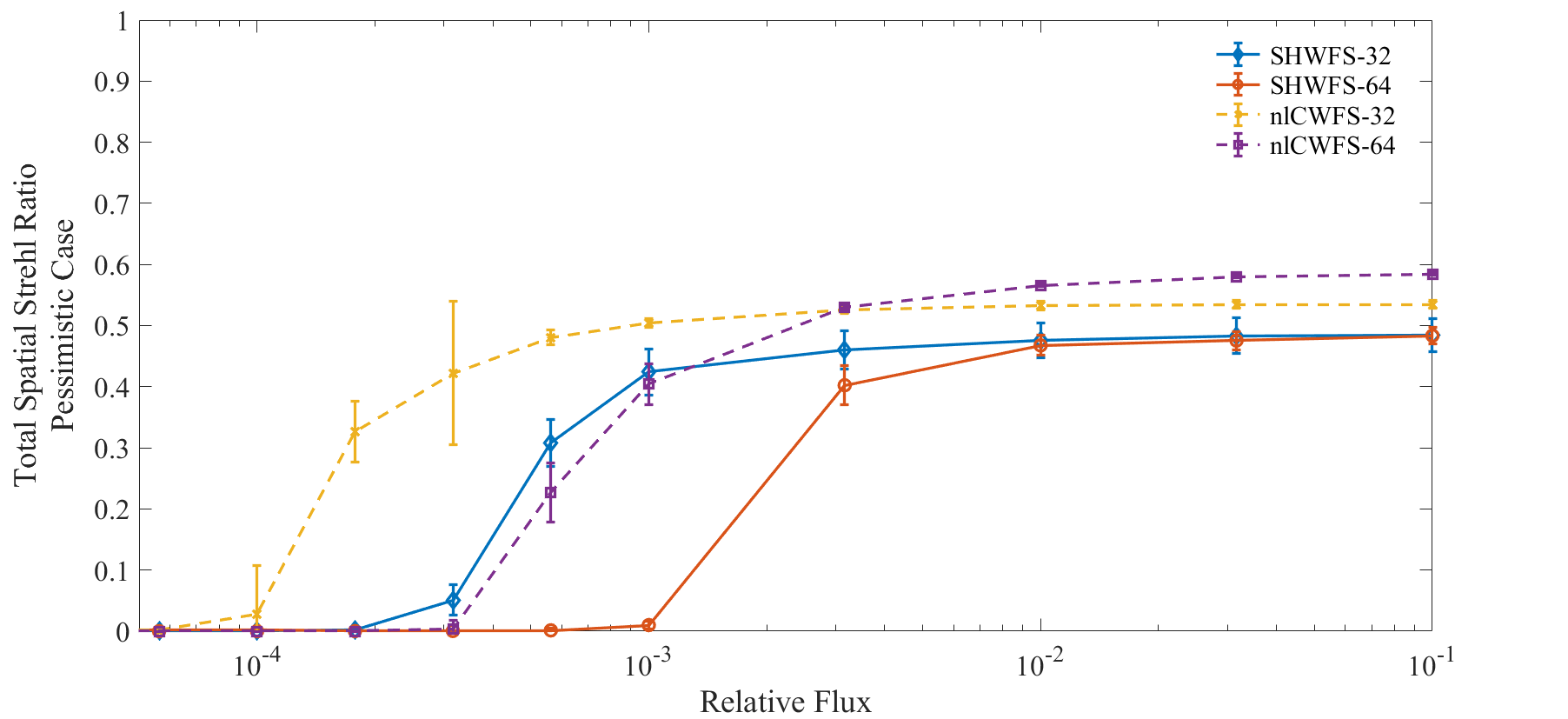}
    \caption{Comparison of the pessimistic SHWFS and nlCWFS spatial Strehl ratio in the presence of diminishing flux.}
    \label{fig:totalWFEestimates_pess_strehl}
\end{figure}

Since the performance of each sensor is uniquely dependent on guide star light intensity, we plot total spatial RMS WFE as a function of $F$ in the optimistic (Fig. \ref{fig:totalWFEestimates_opt}) and pessimistic (Fig. \ref{fig:totalWFEestimates_pess}) cases. Only the RMS WFE incurred through photon noise and read noise is simulated to vary with flux; RMS WFE from other error sources is included through the quadrature addition of the values reported in Tables \ref{tab:WFE_budget_32} and \ref{tab:WFE_budget_64}. Figures \ref{fig:totalWFEestimates_opt_strehl} and \ref{fig:totalWFEestimates_pess_strehl} plot the corresponding optimistic and pessimistic Strehl ratio measurements, which are calculated using the Marechal approximation \cite{Ross:09}.




Figures \ref{fig:totalWFEestimates_opt} and \ref{fig:totalWFEestimates_pess} show that the nlCWFS generates less spatial RMS WFE than the SHWFS, particularly at lower illuminations. At $F>10^{-2.5}$, the nlCWFS maintains a small advantage over the SHWFS, generating up to 10 nm less RMS WFE in the pessimistic case and up to 3 nm less in the optimistic case. These reductions in RMS WFE correspond to improvements of 0.02 and 0.10 in the Strehl ratio in the optimistic and pessimistic cases, respectively. However, the nlCWFS provides a significant performance improvement over the SHWFS at lower illuminations. At $F=10^{-3}$, the nlCWFS generates up to 20 nm less RMS WFE than the SHWFS in the optimistic case and up to 100 nm less in the pessimistic case. Such reductions in RMS WFE correspond to improvements of 0.18 and 0.40 in the Strehl ratio in the optimistic and pessimistic cases, respectively. We conclude that the nlCWFS offers significant performance advantages over the SHWFS, especially for low-flux applications. Further, the nlCWFS's advantages in the spatial domain are expected to grow with increasing turbulence strength due to the nlCWFS's more accurate reconstruction methods. Thus, the nlCWFS is a promising candidate for applications involving deep turbulence and low illumination, which is the subject of a forthcoming article \cite{Potier:23}.






\section{Summary and Concluding Remarks}\label{sec:conclusions}

Although the inherent sensitivity of the SHWFS is significantly lower than the theoretical limit set by photon noise, the simplicity and robustness of its design have offered a pragmatic solution for AO applications for decades. Where there appears to be room for improvement involves science cases that use faint guide stars and operation in sufficiently strong turbulent conditions to cause irradiance fade \cite{Crepp:20}. In this paper, we have developed a spatial domain error budget to help quantify and compare the nlCWFS's and SHWFS's responses to various error sources common to AO systems. 

Our results corroborate the idea that a nlCWFS offers excellent inherent sensitivity (photon noise) and precision (algorithmic errors) as a consequence of its design, which utilizes the wave nature of light instead of geometric optics. Notably, the nlCWFS offers a highly accurate reconstruction of the low spatial frequencies that dominate the atmospheric power spectrum. We also find that the nlCWFS performs well in comparison to the SHWFS when considering the effects of detector vibrations. While differential tip/tilt between propagation channels and other non-common-path aberrations can limit the performance of a nlCWFS when left unchecked, practical solutions can readily be employed to suppress these error sources. 

Read noise was identified by Crepp et al., 2020\cite{Crepp:20} as a potentially serious impediment to the nlCWFS, requiring expensive photon-counting devices. With a focus on this important error budget term, we have developed a simple pedestal subtraction method that appears to ameliorate much of the limitations imposed by read noise. At a relative flux of $F=10^{-3}$, we find that a nlCWFS with 32$\times$32 spatial samples incurs less than 26 nm of RMS WFE using a sCMOS detector with $20e^-$ of read noise (median). For comparison, a 32$\times$32 SHWFS would need to use a quad-cell configuration to offer better resilience to read noise, but this design decision would come at the expense of reconstruction accuracy.


In summary, the spatial domain error budget of the nlCWFS shows promise for the sensor's ability to address the limitations of the SHWFS, but an outstanding question remains regarding its speed (servo lag). Sensors requiring nonlinear reconstruction algorithms have yet to be deployed in the field, due to bottlenecks in data transfer bandwidth and algorithmic latency, but efforts are underway \cite{Ahn:23}. Recent advances in peripheral component interfacing and increases in raw speed and parallelization of modern CPU, GPU, FPGA, and ASIC devices are encouraging. The fact that the nlCWFS reconstructor is based primarily on Fourier transforms suggests that it should be amenable to speed enhancement. How fast a nlCWFS can operate in practice and how the nlCWFS's wavefront error budget compares to the error budgets of other WFS designs, including the Pyramid WFS, are areas of active research that will be addressed in future work.

\subsection*{Disclosures}
The authors declare no conflicts of interest.

\subsection*{Code, Data, and Materials Availability}
Data underlying the results presented in this paper are not publicly available at this time but may be obtained from the authors upon reasonable request.

\subsection*{Acknowledgments}
This research was supported in part by the Air Force Office of Scientific Research (AFOSR) grant number FA9550-22-1-0435. We acknowledge support from Northrop Grumman Space Systems. Sam Potier acknowledges support from the NSF GRFP program (Grant Number: DGE-1841556) and the Notre Dame Arthur J. Schmitt Leadership Fellowship.



\bibliography{report}   
\bibliographystyle{spiejour}   


\vspace{2ex}\noindent\textbf{Sam Potier} is a Physical Scientist working at the US Department of Energy.  He received his PhD from the Department of Physics and Astronomy at the University of Notre Dame in 2023, and a BS in Physics and Mathematics from St. Norbert College in 2017. His current research interests include wavefront sensors, adaptive optics, and the simulation of optics systems.

\vspace{1ex}
\noindent Biographies of the other authors are not available.

\listoftables
\listoffigures

\end{spacing}
\end{document}